\documentclass{aa}
\usepackage{graphicx}

\newcommand{\Omo}{\Omega_{\rm o}}
\newcommand{\lamo}{\lambda_{\rm o}}
\newcommand{\Ho}{H_{\rm o}}
\newcommand{\Snu}{S_\nu}
\newcommand{\yo}{y_{\rm o}}
\newcommand{\inu}{i_\nu}

\newcommand{\hn}{\hat{n}}
\newcommand{\jnu}{j_\nu}
\newcommand{\To}{T_{\rm o}}
\newcommand{\Mgas}{M_{\rm gas}}
\newcommand{\fgas}{f_{\rm gas}}
\newcommand{\fnu}{f_\nu}
\newcommand{\delc}{\delta_c}
\newcommand{\sigo}{\sigma_{\rm o}}
\newcommand{\Dg}{D_{\rm g}}
\newcommand{\Da}{D_{\rm ang}}

\newcommand{\Sobs}{S_\nu^{\rm obs}}
\newcommand{\fwhm}{\theta_{\rm fwhm}}

\newcommand{\msun}{\rm\,M_\odot}
\newcommand{\Mdetect}{M_{\rm det}}
\newcommand{\MdetectUR}{M_{\rm det}^{\rm ur}}
\newcommand{\MdetectOpt}{M_{\rm det}^{\rm opt}}
\newcommand{\MdetectSt}{M_{\rm det}^{\rm st}}

\newcommand{\Ompix}{\Omega_{\rm pix}}
\newcommand{\thetapix}{\theta_{\rm pix}}
\newcommand{\sigpix}{\sigma_{\rm pix}}
\newcommand{\rc}{r_{\rm c}}
\newcommand{\thetac}{\theta_{\rm c}}
\newcommand{\thetavir}{\theta_{\rm vir}}
\newcommand{\Rv}{R_{\rm v}}
\newcommand{\xv}{x_{\rm v}}
\newcommand{\Nmin}{N_{\rm min}}

\newcommand{\sigb}{\sigma_{\rm b}}
\newcommand{\thetadet}{\theta_{\rm det}}
\newcommand{\calG}{{\cal G}}

\newcommand{\tildy}{\hat{y}}

\newcommand{\delNL}{\Delta_{\rm NL}}
\newcommand{\qopt}{q_{\rm opt}}
\newcommand{\qst}{q_{\rm st}}

\begin{document}

   \thesaurus{12.03.1;12.03.3;12.03.4;12.04.2;12.12.1;11.03.1} 
   \title{Sunyaev-Zel'dovich Surveys: Analytic treatment of cluster
        detection}

   \subtitle{}

    \author{J.G.~Bartlett}

   \offprints{J.G. Bartlett}
   \mail{bartlett@ast.obs-mip.fr}
   \institute{Observatoire Midi-Pyr\'en\'ees,
              14, Ave. E. Belin,
              31500 Toulouse,
              FRANCE}
%              Unit\'e associ\'ee au CNRS \\
%               ({\tt http://astro.u-strasbg.fr/Obs.html})
%             $^2$Institut d'Astrophysique Spatiale\\             
%             }

   \date{January 5, 2000}

   \maketitle

   \begin{abstract}
        Thanks to advances in detector technology and observing 
techniques, true Sunyaev--Zel'dovich (SZ) surveys will soon become a 
reality.  This opens up a new window into the Universe, in many
ways analogous to the X--ray band and inherently well--adapted
to reaching high redshifts.  
I discuss the nature, abundance and
redshift distributions of objects detectable in ground--based
searches with state--of--the--art
technology.  An advantage of the SZ
approach is that the total SZ
flux density depends only on the
thermal energy of the intracluster gas
and not on its spatial or temperature structure,
in contrast to the X--ray luminosity.
Because ground--based surveys will be characterized
by arcminute angular resolution, they will resolve
a large fraction of the cluster population.
I quantify the resulting consequences for
the cluster selection function; these include
less efficient cluster detection compared
to idealized point sources and corresponding
steeper integrated source counts.  This
implies, contrary to expectations based
on a point source approximation, that deep
surveys are better than wide ones in terms of 
maximizing the number of detected objects.
At a given flux density sensitivity and angular
resolution, searches at millimeter wavelengths
(bolometers) 
are more efficient than centimeter searches
(radio), due to the 
form of the SZ spectrum.  Possible ground--based
surveys could discover up to $\sim 100$ clusters
per square degree at a wavelength of 2 mm 
and $\sim 10$/sq. deg. at 1 cm, modeling
clusters as a simple self--similar population.

   \end{abstract}

%________________________________________________________________

\section{Introduction}

        Cosmologists have long appreciated the value of the
Universe's biggest objects, galaxy clusters.  Besides being
a collection of galaxies well suited for studies of galaxy
formation, studies focussed on the global properties of clusters
provide information on the nature of dark matter; the
relative proportions of hot gas, dark matter and stars;
and on scenarios of structure formation,
including constraints on the universal density parameter, $\Omo$.
One example of the latter that comes to mind in anticipation of
observations with the new generation of X--ray satellites,
Chandra and XMM, is the use of the redshift evolution of
the cluster abundance to constrain $\Omo$ (Oukbir \& Blanchard 1992,
1997; Bartlett 1997; Henry 1997; Bahcall \& Fan 1998; 
Borgani et al. 1999; Eke et al. 1998; Viana \& Liddle 1999a); 
another is the now classic cluster baryon fraction test
(White et al. 1993).

        While clusters have been extensively studied in the optical
and X--ray bands, observations based on weak gravitation lensing
and the Sunyaev--Zel'dovich (SZ) effect (Sunyaev \& Zel'dovich 1972) 
are just coming to fruition.
In the case of SZ observations, important samples
consisting of several tens of clusters pre--selected
in the X--ray are beginning to permit cosmologists to
capitalize on the potential of combined SZ/X--ray observations
(Carlstrom et al. 1996, 1999).
Full maturity of the field will be heralded by the
realization of purely SZ--based sky surveys.
In what we might refer to as the ``SZ--band'',
one can then imagine performing cluster science 
analogous to what is now done in the X--ray,
e.g., the construction of cluster counts, redshift
distributions, luminosity functions, etc., 
all viewed via the unique characteristics 
of the the SZ effect.  For example, several authors
have emphasized the advantages of the SZ effect, over similar
X--ray based efforts, to constrain
$\Omo$ via the cluster redshift distribution, as well as 
to study cluster physics out
to very large redshifts (provided the clusters are
out there, the very question of $\Omo$ itself) 
(Korolyov et al. 1986; Bond \& Meyers 1991; Bartlett \& Silk 1994;
Markevitch et al. 1994; Barbosa et al. 1996; Eke et al. 1996; 
Colafrancesco et al. 1997; Holder et al. 1999).   
Such pure SZ surveys will be performed: the Planck Surveyor 
will supply an almost full--sky catalog of several
thousand clusters detected uniquely by their SZ signal;
and advances in both detector technology
and observing techniques now offer the 
exciting prospect of performing purely SZ--based surveys 
from the ground, with both large format bolometer
arrays and dedicated interferometers.  

        I discuss in this {\em paper} some aspects of the science 
accessible to pure SZ surveys by examining the nature 
of their cluster selection.
Because of the close analogy with X--ray studies, it 
is useful for this purpose
to compare and contrast SZ--based cluster searches to those
based on X--ray observations.
The redshift independence of the surface brightness of a cluster
(of given properties) means that SZ cluster detection is inherently
more efficient than X--ray detection at 
finding high redshift objects.
Equally important is that although the SZ effect 
and X--rays both ``see'' the hot 
intracluster medium (ICM), they do
so in significantly different ways.  In particular, the
well--known fact that the SZ effect scales as the gas pressure
implies that the flux density, $\Snu$, is simply 
proportional to the {\em total thermal 
energy} of the gas.  This makes modeling especially simple, for
this quantity depends only on the {\em total gas mass} and the 
{\em effectiveness
of gas heating} during collapse, in stark contrast to the
X--ray emission that depends also on the density and temperature
distribution of the gas.  This simplicity is an advantage because
any theoretical interpretation of survey results
requires an adequately modeled relation between
the observable and the theoretically relevant quantity
of cluster mass.    
%SZ surveys
%should also reach the low--end of the mass function inaccessible 
%to X--rays, which have great difficulty finding objects with
%temperatures lower than $\sim 1$ keV, due to the emitting/absorbing
%screen presented by our own Galaxy.  Not suffering from these foreground
%effects, the SZ--band opens the window on very low mass groups,
%such as the Local Group, detecting them in an unambiguous manner
%%through the presence of shock--heated gas, rather than less sure
%group finding algorithms applied to redshift surveys.

        These remarks concern essentially the physics
of the `emission' mechanism itself.  Of equal 
relavence is the nature of object selection imposed
by the eventual detection algorithm used to extract
sources from a set of observations (a map); 
and this in turn depends crucially, as for any survey,
on the particular combination of {\em sensitivity} and 
{\em angular resolution} 
of the observations.  The objects detected by 
Planck will not be the same as those selected
by ground--based surveys, and
the final catalogs should be viewed as complementary.
Planck will produce a shallow 
($\sim$ tens of mJy) large--area survey, while
the ground--based instruments will perform deeper surveys
($< 1$ mJy) over smaller sky areas (several
square degrees).  Most clusters remain unresolved at the
Planck resolution of $\sim 5-10$ arcmins, and 
this characterizes the kinds of objects accessible
to this survey, e.g., the counts and the 
redshift distributions.  The higher angular
resolution of future ground--based instruments
(on the order of an arcmin) will 
resolve many clusters and impose different 
selection criteria that will define the counts and redshift
distributions of the final catalog.  

        This is a
central issue of the present study were, motivated
by the possibility of ground--based surveys, I examine
the detection of {\em resolved} clusters.
While the detection of unresolved sources is principally 
dependent on observational sensitivity, and
the final selection is more or less one of apparent
flux -- $\Snu \sim \thetac^2\inu$ --  the detection of resolved sources
is a more complicated cuisine involving individually
the characteristic source size, $\thetac$, and 
surface brightness, $\inu$.  
The specific goal of the present work is to 
quantify in terms of observing parameters
the abundance, masses and redshifts of
clusters detectable by ground--based surveys, with the particular
aims of understanding optimal object extraction
and the accessible science. 
For example, one of the key questions facing any survey is one
of observing strategy: given a fixed, total amount of
observing time, should one ``go deep'', with long 
integrations on a few fields, or instead ``go wide'', covering
more fields to higher sensitivity.  If one is out to 
maximize the number of detected objects, the answer depends on the
slope of the counts.  One gains by going deeper if the integrated 
counts are steeper than $\Snu^{-2}$, assuming that 
noise diminishes as $1/\sqrt{t}$; otherwise, a larger area yields
more objects.  

%For unresolved SZ sources, a situation 
%more appropriate to the Planck Surveyor, the predicted integrated 
%source counts are roughly Euclidean -- $N(>S)\sim S^{-3/2}$ -- and
%the slope never exceeds the critical value of $-2$.  On the
%other hand, and as developed below, resolved SZ counts are
%steeper, suggesting that deep surveys may be the
%best strategy: besides optimizing the total number of 
%detections, they probe lower masses and farther in redshift.
%This situation is more applicable to the ground--based
%facilities which will soon be capable of surveying.

        The cluster selection criteria of a survey may be
compactly summarized by a minimum detectable mass as a 
function of redshift -- $\Mdetect(z)$.  Together with
a suitable mass function (we shall use the formalism
of Press and Schechter 1974), this quantity 
determines both the source counts and redshift 
distributions of the final source catalog.
Thus, in very concrete terms, we must
examine $\Mdetect(z)$ and understand 
the influence of the observationally imposed 
restrictions on $\thetac$ and $\inu$. 
Given a set of observations, i.e., a {\em map}, one could 
imagine many different algorithms to extract astrophysical sources,
and $\Mdetect(z)$ will depend upon this choice.
There is in principle an optimal method, 
one which preserves signal--to--noise over the 
entire range of source surface brightness and size.  
It is characterized by a decreasing surface brightness
limit with object size -- the greater number of object
pixels permits lower surface brightness detections. 
This algorithm is difficult to apply in practice,
and more standard approaches search instead for
a minimum number of connected pixels above a 
preset threshold, thereby establishing a fixed 
cut on source surface brightness.  Detection 
signal--to--noise is no longer constant, rather 
increasing with $\thetac$, and these methods loose
large, and in--principle detectable, low surface 
brightness objects.  All of this will be reflected
in the resulting functions $\Mdetect(z)$.

%We will discuss these issues in detail by
%first examining the optimal, constant
%signal--to--noise extraction, in order to understand
%the intrinsic differences of resolved detection 
%compared to point source extraction.  Then 
%we address the standard algorithm, comparing
%it in turn to the optimal method.  

        Throughout the discussion, we will be guided 
by the characteristics of two potential types of ground--based
instruments: large format bolometer arrays, epitomized by
BOLOCAM (Glenn et al. 
1998\footnote{\tt http://phobos.caltech.edu/~lgg/bolocam/bolocam.html }), 
and interferometer arrays
optimized for SZ observations, as suggested by Carlstrom et al. 
(1999)\footnote{While writing, I became aware of another project --
the Arcminute MicroKelvin Imager.  See Kneissl R. 2000, 
astro-ph/0001106}.
BOLOCAM is a 151--element bolometer array under construction
at Caltech for operation in three bands -- 2.1~mm, 1.38~mm
(the null of the thermal SZ effect) and 850~$\mu$m.  
At the Caltech Submillimeter Observatory, it is expected
that the array will be diffraction limited to $\sim$ arcminute
resolution, or better, and limited in sensitivity 
by atmospheric emission (rather than detector noise).
%to $\sim 35$ mJy/$\sqrt{\rm Hz}$.  
With its $9$--arcmin field--of--view, one could
imagine surveying a square degree to sub--mJy 
sensitivity in these bands.
%This means, for 
%example, that with a one hour exposure, one could
%reach sub--mJy sensitivity/pixel.  With its 
%$9$--arcmin field--of--view, a full square degree
%could be imaged to this sensitivity in roughly 60 hours 
%of observation time.
Carlstrom et al. (1999) have recently expounded the
virtues of interferometric techniques
using telescope arrays specifically designed for SZ 
observations.  They have proposed the construction of
such an array, operating at a wavelength of 1 cm, 
and estimated that it would be capable,
in the course of one year of dedicated observations,
of covering $\sim 10$ square degrees to a limiting sensitivity
of $\sim 0.3$ mJy at arcminute resolution.  In summary, then,
we are interested in considering SZ observations
at arcminute angular resolution and to sub--mJy
sensitivity at both centimeter and millimeter wavelengths.

        The {\em paper} is organized as follows:
a rapid review of the SZ effect is given in the next
section, followed by a discussion of the unique
aspects of SZ cluster detection.  Section 3 
details the cluster population model employed,
based on the Press--Schechter (Press \& Schechter, 1974)
mass function and the isothermal $\beta$--model.  
Since we shall focus on issues of 
cluster selection as imposed by survey
parameters, the cluster model
will be restricted to the simple example of a 
self--similar population.  The next
section (Section 4) introduces the 
principal figures (Figures 1,2 and 3) 
of the present work by consideration of unresolved cluster
detection; this case will also be used
as a benchmark against which to examine
the effects of resolved detection.
Section 5 then develops the principle
themes of resolved SZ cluster detection,
starting with consideration of the optimal,
constant signal--to--noise method, and 
followed by detailed study of cluster detection
based on the standard algorithm.  A final
discussion (Section 6) then more closely
examines the number of detections to be
expected from ground--based surveys
and gives a non--exhaustive list
of some important issues still to be
treated.  Section 7 concludes.

     Key results will be the $\Mdetect(z)$ curves
presented in Figure 1, quantifying the nature of 
SZ detected clusters, and the conclusion that
resolved source counts are lower and steeper than expectations
based on simple unresolved source count
calculations, Figure 2.  To the point, 
the latter implies that surveys at arcminute resolution
gain objects with an observing strategy of ``going deep''. 
The cosmological
density parameter is denoted by $\Omo\equiv 8\pi G\rho/3\Ho^2$,
the vacuum density parameter by $\lamo\equiv \Lambda/3\Ho^2$
and the Hubble constant by $\Ho\equiv h 100$ km/s/Mpc; 
unless otherwise indicated, $h=1/2$ and $\lamo=0$.

\section{The Particular Value of the SZ Effect}

        We begin by establishing our notation in recalling the 
basic formulas of the SZ effect.  The change in surface 
brightness relative to the unperturbed cosmic microwave 
background (CMB), caused by inverse Compton scattering in the
hot ICM, is expressed as
\begin{equation}
\inu(\vec{\theta}) = y(\vec{\theta})\jnu(x)
\end{equation}
where $x\equiv h_p\nu/k\To$ is a dimensionless frequency
expressed in terms of the energy of the unperturbed CMB Planck
spectrum at $\To=2.725\;$ K (Mather et al. \cite{cmb:temp2}).
%$\To=2.728\;$ K (Fixsen et al. \cite{cmb:temp1}).
The spectral shape is embodied in the function $\jnu$,
\begin{eqnarray}
\label{jnu}
\nonumber
\jnu(x) & = & 2\frac{(k\To)^3}{(h_pc)^2} \frac{x^4\mbox{e}^x}{(\mbox{e}^x-1)^2}
        \left[\frac{x}{\tanh(x/2)} - 4\right] \\
& \equiv & 2\frac{(k\To)^3}{(h_pc)^2} f_\nu \\
\nonumber 
& = & (2.28\times 10^4 {\rm mJy/arcmin}^2) f_\nu
\end{eqnarray}
while the amplitude is given by the Compton $y$--parameter
\begin{equation}
\label{comptony}
y \equiv \int \mbox{d}l \frac{kT}{m_ec^2} n_e \sigma_T
\end{equation}
an integral of the {\em pressure} along the line--of--sight at
position $\vec{\theta}$ relative to the cluster center.
Here, $T$ is the temperature of the ICM (really, the electrons), 
$m_e$ is the electron rest mass, $n_e$ the ICM electron density, 
and $\sigma_T$ is the Thompson cross section.
Planck's constant is written in these expressions as
$h_p$, the speed of light in vacuum as $c$, and
Boltzmann's constant as $k$.  These formulae apply in
the non--relativistic limit of low electron (and photon) 
energies; relativistic extensions have recently been 
made by several authors (e.g., Rephaeli 1995; Stebbins 1997;
Challinor \& Lasenby 1998; Itoh et al 1998; Pointecouteau et al. 1998;
Sazonov \& Sunyaev 1998).
%Notice the introduction
%of the dimensionless spectral function $f_\nu$.
The spectral shape of the distortion is unique, becoming negative
at wavelengths larger than $\sim 1.4$ mm (relative to ``blank'' sky) and
positive a shorter wavelengths.  This offers a way of clearly 
separating the effect from other astrophysical emissions.

        All of the physics is in the Compton $y$--parameter,
an apparently innocuous--looking expression.  In fact, 
it holds the key to all of the pleasing aspects of the 
SZ mechanism.  First of all, the conspicuous absence
of an explicit redshift dependence  
is the well--known result that the SZ surface brightness 
is redshift--independent, determined only by cluster properties.  
This should be contrasted to other emission mechanisms which
all experience ``cosmic dimming'' [$\iota \propto (1+z)^{-4}$] 
due to the expansion of the Universe.  
This is countered in the SZ effect by the increasing energy density 
towards higher $z$ of the CMB, the source of photons for the effect.

        Another very important aspect of the SZ mechanism 
resides in the fact that its amplitude is proportional to the pressure, 
or {\em thermal energy}, of the ICM.  This appears most clearly
when we consider the total flux density from a cluster, found
by integrating the surface brightness over the cluster face:
\begin{eqnarray}
\label{eq:fluxdens}
\nonumber
\Snu(x,M,z) & = & \jnu(x) D_a^{-2}(z) \int \mbox{d}V \frac{kT(M,z)}{m_ec^2} 
        n_e(M,z) \sigma_T \\
& \propto & \Mgas <T>
\end{eqnarray}
The integral is over the entire virial volume of the cluster.
In this expression, $D_a(z)$ is the angular--size distance
in a Friedmann--Robertson--Walker metric -- 
\begin{eqnarray}
\label{Dang} 
\nonumber
\Da(z) = 2c\Ho^{-1} \left[\frac{\Omo z + (\Omo - 2)(\sqrt{1+\Omo z} - 1)}
        {\Omo^2 (1+z)^2}\right] \\
\; = c\Ho^{-1} D(z)
\end{eqnarray}
where I introduce the dimensionless quantity $D(z)$.
We see clearly that the final result is simply
{\bf proportional to the total thermal energy of the 
ICM}, $\int dV n T$.  This is extremely important, 
because it means that the SZ flux density is {\em insensitive
(strictly speaking, completely so for the total flux density
and for fixed thermal energy) to 
either the spatial distribution of the ICM or its temperature 
structure}, making
modeling much simpler than in the case of X--ray emission.
Consider that in X--ray modeling one prefers the
X--ray temperature over luminosity as a more robust 
indicator of cluster mass, but even the temperature 
has some sensitivity to the
gas distribution, because it is all the same an emission weighted
temperature that is actually observed.  
%In addition, 
%recent work by Mathieson et al. (1999) suggests 
%that the X--ray temperature, which is in practice found by 
%spectral fitting, can also 
%be affected by the line emission of
%lower temperature gas in clusters, and this
%will depend on the such things as the ICM metallicity.
We would expect the temperature
appearing in the second line of Eq. (\ref{eq:fluxdens}),
which is the {\em true} mean electron energy,
to demonstrate an even better correlation with virial 
mass than the observed X--ray temperature.
%In contrast, the mean temperature appearing in the
%second line of Eq. (\ref{eq:fluxdens}) is the true mean
%electron energy.  From the which we would expect to demonstrate an 
%even better correlation with virial mass than the 
%observed X--ray temperature, which 
%in numerical simulations already
%demonstrates a good correlation with cluster mass (REF).  
Simple scaling arguments lead one to believe that this 
correlation should be 
$T\sim T_{\rm virial}\sim M^{2/3} (1+z)$, from which we deduce
\begin{equation}\label{eq:SZMcorr}
\Snu \sim \fgas(M,z) M^{5/3} (1+z) D^{-2}(z)
\end{equation}
where $\fgas$ is the gas mass fraction contributed by the ICM
to the total cluster mass.

        The SZ mechanism therefore conveniently 
reduces all the potential complexity of the ICM 
to just its total thermal energy, $\propto \fgas <T>$.  
This quantity may nevertheless be influenced by several 
factors.  For example, the gas mass fraction 
in Eq. (\ref{eq:SZMcorr}) has carefully been written 
as a general function of both mass and redshift. 
In simulations this quantity is most often 
constant, the majority of gas being primordial and simply
falling into the cluster at formation.  One could imagine
other possibilities (e.g., Bartlett \& Silk 1994; 
Colafrancesco \& Vittorio 1994) that would lead to a more important
dependence on either mass or redshift, although metallicity
arguments seem to require that most of the gas be
primordial, at least in the more massive systems 
(Metzler \& Evrard 1994; Elbaz et al. 1995).
While it appears from numerical studies that shocking during
cluster formation efficiently heats the ICM to $\sim 80$\%--
$100$\% of the
virial temperature (Metzler \& Evrard 1994; Bryan \& Norman 1998), 
additional sources of heating could
in principle change the temperature of the gas relative to 
that of the potential, i.e., $T\ne T_{\rm virial}$.  
Such heating may not always 
produce the most obvious effects -- remember that it is
the total thermal energy of the gas that counts, and 
understanding the change of this quantity with heating 
in a gravitational potential requires careful modeling.
Although models studied so far do not lead to 
a strong effect (Metzler \& Evrard 1994), we shall at times 
be discussing rather
low mass systems, for which these effects are poorly
understood theoretically and observationally.   Finally, the exact 
form of the virial temperature--mass relation depends in part on the
dark matter profile of the collapsing proto--cluster;
once again, numerical experiments seem to indicate that
this does not change too much, i.e., one finds a good
T--M relation with rather small scatter 
(Evrard et al. 1996; Bryan \& Norman 1998).  
Putting all of 
this together, a relation of the form (\ref{eq:SZMcorr}) 
between the observable, $\Snu$, and cluster mass
appears quite reasonable and rather robust; and in any case, 
the modeling uncertainties are always easier to understand than
in the case of X--rays, due to the all important insensitivity of 
the SZ flux density to spatial/temperature structure of the ICM.

        The conclusion is that the SZ flux
density should be a very good {\em halo mass detector},
in principle sensitive to all halos with significant
amounts of hot gas and over a large range of redshifts.
%We may also note in this light the existence of very low mass
%groups, i.e., halos larger than galaxies but smaller
%than the lowest temperature X--ray detected groups, say
%with masses in the range $few\times 10^{12}-10^{14}\; \msun$,
%such as the Local Group.  These objects are extremely difficult,
%if not impossible, to detect in X--rays because the
%optical depth of our Galaxy becomes important below
%energies of $\sim 1$ keV.  In this way, the SZ effect could
%open the window on a new mass range of objects,
%those in the interesting region between galaxies and clusters,
%where gas cooling, star formation and feedback effects play
%important roles.  This does make the modeling
%more uncertain in this mass range (e.g., see above),
%but this is precisely because we know so little about it
%at the present time.
All of these remarks concern
to a large extent the total SZ flux density of 
Eq. (\ref{eq:fluxdens}), and therefore apply primarily
to situations where the clusters are unresolved.
It is still true that, even when a cluster is 
resolved, the SZ signal is proportional to the
total thermal energy of the gas, but now only
of that portion contained within the column 
defined by the beam.  
After first outlining the cluster population model
employed, we shall tackle in detail the
additional complexities introduced by 
{\em resolved} cluster observations.
 
\section{Modeling the Source Population}

        The central ingredient of a model for the
cluster population and its evolution is the
mass function, $n(M,z)$, which gives the number density
of collapsed, virialized objects as a function
of mass and redshift.  The exact form of this 
function depends on the statistical properties
of the primordial density fluctuations.  
For Inflationary--type scenarios, in which these 
fluctuations are Gaussian, a reasonable expression 
for the mass function appears to be 
the Press--Schechter formula (Press \& Schechter 1974)
\begin{equation}
\label{eq:psfunc}
n(M,z) dM = \sqrt{\frac{2}{\pi}} \frac{<\rho>}{M} \nu(M,z) 
        \left| \frac{\mbox{d}\ln \sigma(M)}{\mbox{d}\ln M} \right| 
\mbox{e}^{-\nu^2/2} \frac{dM}{M}
\end{equation}
The quantity $\langle\rho\rangle$ represents 
the {\em comoving} cosmic mass density
and $\nu(M,z)\equiv \delc(z)/\sigma(M,z)$, with
$\delc$ equal to the critical {\em linear} over--density
required for collapse and $\sigma(M,z)$ the 
amplitude of the density perturbations on a 
mass scale $M$ at redshift $z$.  Numerical studies
ascribe rather remarkable accuracy to the simple expression  
of Eq. (\ref{eq:psfunc}) (Lacey \& Cole 1994;
Eke et al. 1996; Borgani et al. 1999), and we shall adopt it
in the following.  
More explicitly, $\delc(z,\Omo,\lamo)$ and $\sigma(M,z)=\sigo(M)\times
(\Dg(z)/\Dg(0))$, with $\Dg(z,\Omo,\lamo)$ being the linear growth
factor.  It is essentially through $\Dg$ that the
dependence on cosmology ($\Omo$, $\lamo$) enters the 
mass function, with $\Omo$ being the more important
of the two as the dependence on $\lamo$ is relatively
weak (see, e.g., Bartlett \cite{casa} for a detailed
discussion).  This dependence on $\Omo$ in the exponent
means that the cluster abundance as a function of 
redshift is a very sensitive probe of the density parameter
(e.g., Oukbir \& Blanchard 1992, 1997), and is 
the motivation for many efforts in all wavebands to
find clusters at high redshifts.
As emphasized by several 
authors (Barbosa et al. 1996; Eke et al. 1996; Colafrancesco et al. 1997; 
Bartlett et al. 1998; Holder \& Carlstrom 1999; Mohr et al. 1999), 
the SZ effect
is particularly well positioned in this arena (see also below).  

        It is clear that the important theoretical variables
are cluster mass and redshift.  Although redshift is 
directly measurable, the mass appearing in Eq. (\ref{eq:psfunc})
must be translated into an observational quantity suitable for 
the type of observations under consideration.
As mentioned above, one of the pleasant features of the SZ
effect is the simplicity of this relation.  
Using the simulations of Evrard 
et al. (1996) to normalize the $T-M$ relation, 
we can quantitatively express the {\em total SZ flux density} 
of a cluster (e.g., Eqs. \ref{eq:fluxdens} \& \ref{eq:SZMcorr}) as 
\begin{eqnarray}
\label{eq:SZfluxdens}
\nonumber
S_{\nu} = (34\,\mbox{\rm mJy}\, h^{8/3}) f_\nu(x) \fgas 
        \Omega_o^{1/3}
        \left[\frac{\delNL(z)}{178}\right]^{1/3}M_{15}^{5/3} \\
        (1+z) D^{-2}(z)
\end{eqnarray}
where the mass $M_{15}\equiv M/10^{15}\;
\msun$ refers to the cluster virial mass and
$\fgas$ is possibly a function of both mass and redshift
(see also Barbosa et al. 1996, but note that the 
definition there of $D(z)$ differs by a factor of $2$).
Evrard (1997) finds $\fgas=0.06\; h^{-1.5}$,
while Mohr et al. (1998) find marginal evidence
for a decrease in lower mass systems (see also Carlstrom et al.
1999 for recent work based on SZ images); 
% Mohr: 0.075h^-1.5 in high mass systems
% Grego: 0.085+0.011-0.015h^-1
there is little information on any possible evolution with redshift at
present.  Other quantities appearing in this equation are the 
mean density contrast for virialization, $\delNL(z,\Omo,\lamo)$
($=178$ for $\Omo=1$, $\lamo=0$), and the dimensionless functions
$f_\nu$ and $D(z)$ introduced in Eqs. (\ref{jnu}) and
(\ref{Dang}).  

\begin{figure}
\resizebox{\hsize}{!}{\includegraphics{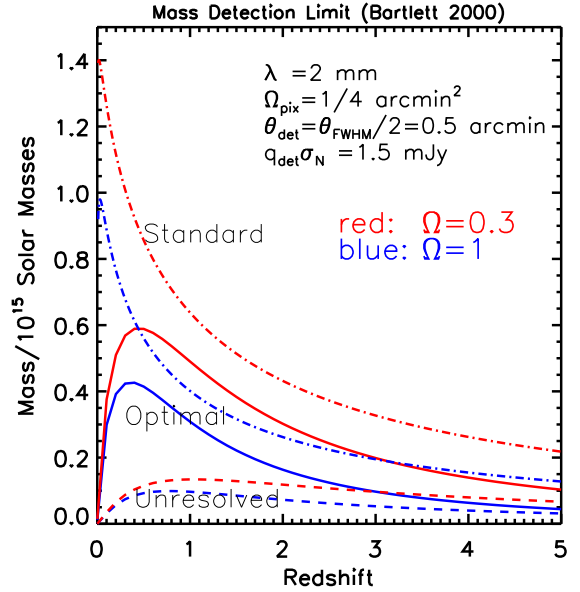}}
\caption{{\bf a)}Detection mass as a function of redshift
for unresolved (dashed lines), optimal resolved (solid lines)
and standard resolved (dot--dashed lines) detection
satisfying $q_{\rm det}\sigpix=1.5$ mJy 
at a wavelength of $2$ mm.  In the unresolved case,
this simply corresponds to the limiting total flux
density.  For optimal resolved detection, $q_{\rm det}$
refers to $\qopt$ in Eq. (\ref{eq:OptMdet}), while
for standard resolved detection it refers to 
$\qst$ of Eq. (\ref{eq:thetadet_dimless}).  The pixel
size has been taken to be $\fwhm/2$, and for 
the standard routine a detection angle
$\thetadet=1/2\fwhm$ has been assumed, as indicated.
In all cases these parameters correspond to $3\sigma$
detections (see text for more detail).
The upper (red) curves in each case correspond to the 
open model with $\Omega=0.3$.}
\end{figure}
\setcounter{figure}{0}
\begin{figure}
\resizebox{\hsize}{!}{\includegraphics{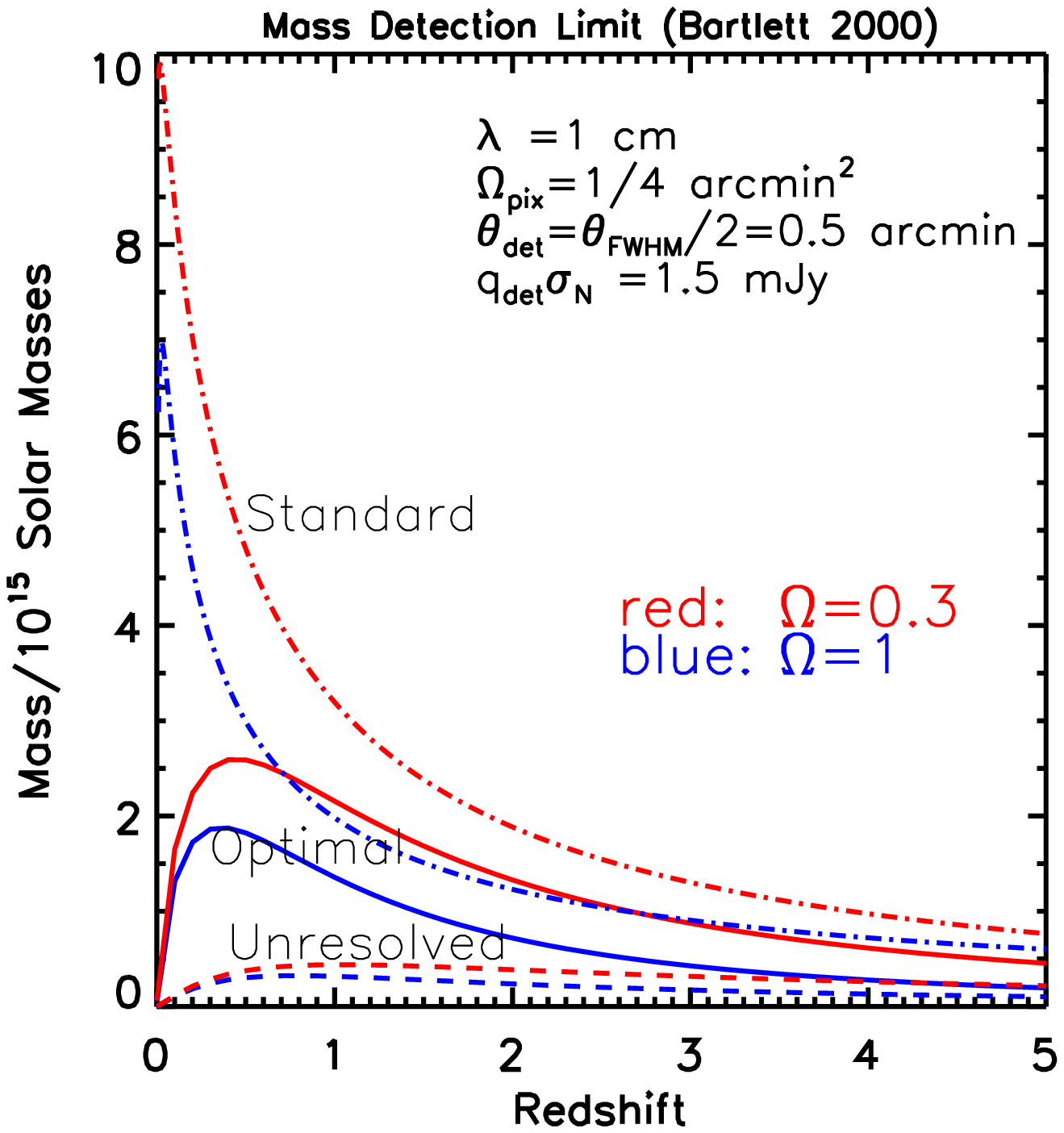}}
\caption{{\bf b)}Detection mass at $1$ cm for the listed
parameters.  The curves are labeled as in the previous figure, but
note the change in scale along the $y$--axis.}
\end{figure}
\setcounter{figure}{0}
\begin{figure}
\resizebox{\hsize}{!}{\includegraphics{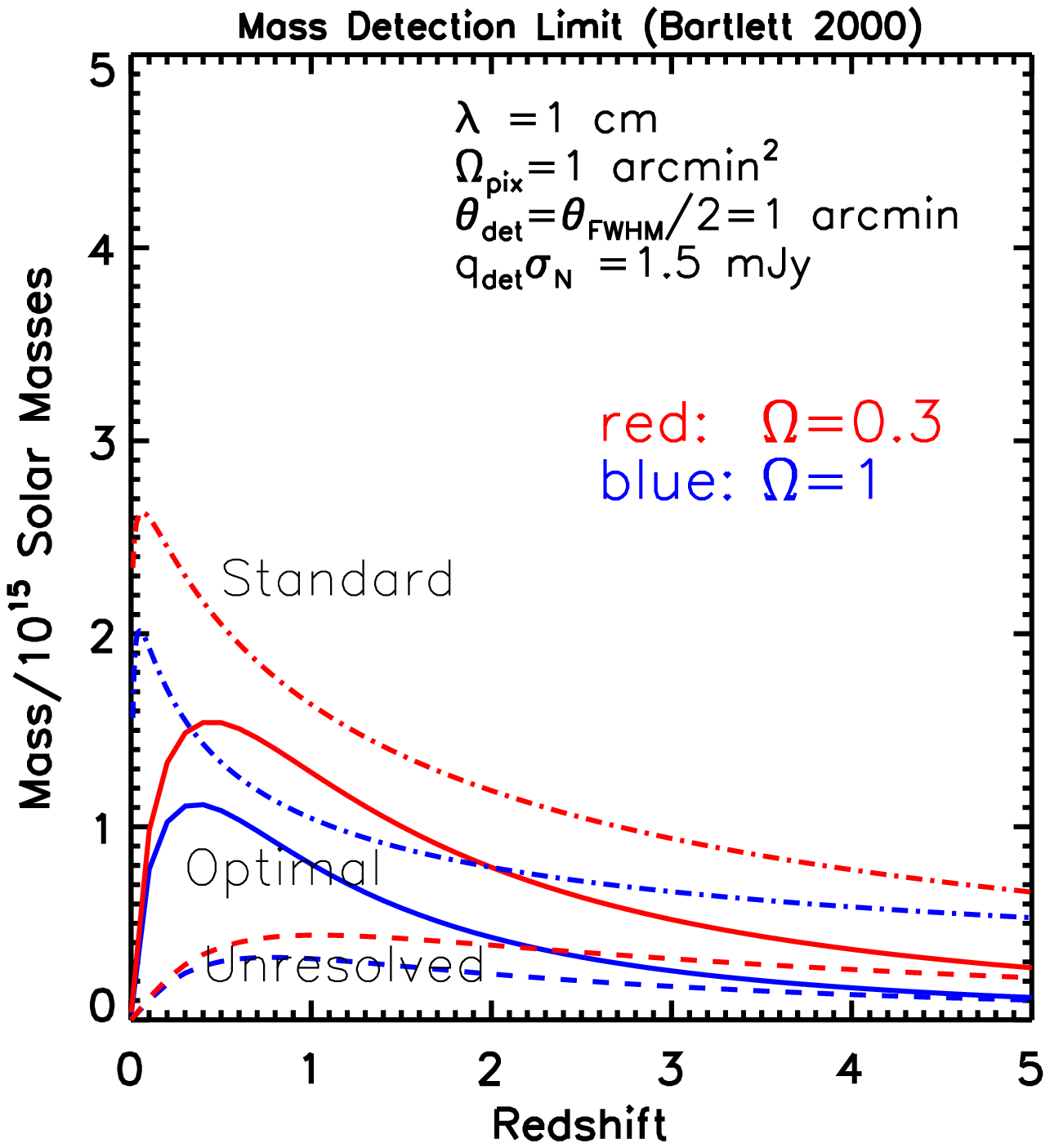}}
\caption{{\bf c)}Detection mass at $1$ cm and for $\Ompix=1$ 
arcmin$^2$ ($\fwhm=2$ arcmins).
The curves are labeled as in the previous figures.  Relative
to Figure 1b, the lower resolution results in smaller detection 
masses (note again the change in ordinate scale).  The
unresolved detection curves are unaffected by the change in resolution.}
\end{figure}

     Observations for which clusters are unresolved 
measure this total flux density, and therefore
this is all that is needed in order to calculate
the unresolved source counts, as we will do in the
next section.  For resolved sources, on the other hand,
the detection criteria are more
complicated.  Contrary to the point source limit,
the details of the cluster SZ profile now
play an important role.  
I will employ a simple isothermal $\beta$--model
to describe this profile: 
\begin{equation}\label{eq:bmodel}
i_\nu(\vec{\theta}) = \frac{\yo\jnu(x)}{(1+\theta^2/\thetac^2)^\alpha}
\end{equation}
The exponent $\alpha=0.5(3\beta-1)$,
where $\beta$ is the exponent of 
the three--dimensional ICM density profile: 
\mbox{$n \propto (1+r^2/\rc^2)^{-3\beta/2}$},
$\rc$ being the physical core radius.
Local X--ray observations indicate
that $\beta\sim 2/3$, a value I 
adopt throughout for the calculations.
In this case, $\alpha=1/2$, a rather
significant value, as will be discussed 
shortly.  This profile will be assumed to hold
out to the virial radius, $\Rv$, of the
cluster.  

     The $\beta$--profile of Eq.(\ref{eq:bmodel}) is
empirically described by $\yo$, a sort of central
surface brightness (actually, it is $\yo\jnu$ that has
units of surface brightness, but it is simpler
to work with $\yo$), and $\thetac$.  In these
terms, there is nothing specific to the SZ effect.
The physics of the SZ effect appears only when
we make the connection between these empirical
parameters and the theoretically interesting
ones, namely, mass and redshift, via relations of 
the kind $\yo(M,z)$ and $\thetac(M,z)$.
As our principle goal in this work is to understand
the selection effects of resolved SZ cluster detection, 
the model for cluster evolution will be kept simple:
a constant gas mass fraction, $\fgas = 0.06\; h^{-1.5}$
(Evrard 1997), over
cluster mass and redshift, and a core
radius scaling with the virial radius $\Rv$, i.e.,
$\xv\equiv \Rv/\rc = const$. {\bf Unless otherwise specified,
this constant will be given a value of 10}.  
One deduces from simple scaling arguments that 
\begin{displaymath}
\Rv = (1.69 h^{-2/3} {\rm\ Mpc})\; M_{15}^{1/3} (1+z)^{-1}
        \Omega_o^{-1/3} \left(\frac{178}{\delNL(z)}\right)^{1/3}
\end{displaymath}
where the normalization is taken from the spherical
collapse model.  This scaling relation is about as robust as 
the relation for cluster temperature; in fact, the two
are essentially the same, since $T\sim M/\Rv$.  Some 
dependence of the normalization on mass and redshift could appear if
the density profile around a peak forming a cluster
changed significantly with these two quantities.
In the following, we shall ignore this possibility, which
numerical simulations seem to indicate is a small effect
in any case.  This then fixes the relation 
\begin{equation}\label{eq:thetac}
\rc(M,z) = \Rv(M,z)/\xv  
\end{equation}

     For the axially symmetric surface brightness of
Eq. (\ref{eq:bmodel}), the integral defining the 
total SZ flux density may be written
\begin{eqnarray}
\nonumber
\Snu(M,z) & = & \jnu 2\pi \int d\theta\theta\; y(\theta)\\
\nonumber
& = & 2\pi\jnu \yo(M,z)\thetac^2(M,z)\left(\sqrt{1+\xv^2}-1\right)
\end{eqnarray}
Using Eq. (\ref{eq:SZfluxdens}) for $\Snu(M,z)$ in 
this expression, we deduce
\begin{eqnarray}\label{eq:yo}
\nonumber
\yo(M,z) & = & (6.40\times 10^{-5} h^2) \fgas\Omo
            \left(\frac{\delNL(z)}{178}\right) \\
& &            M_{15} (1+z)^3 
            \left(\frac{\xv^2}{\sqrt{1+\xv^2}-1}\right)
\end{eqnarray}
Together with the $\beta$--profile (Eq. \ref{eq:bmodel}), 
Eqs. (\ref{eq:thetac}) and (\ref{eq:yo}) define our
cluster evolution model.  As mentioned, it is self--similar,
and we see the expected scaling \mbox{$\rc\sim M^{1/3}/(1+z)^{-1}$}
and \mbox{$\yo\sim M(1+z)^3$}.  This is most probably
an oversimplified description of the actual
cluster population, but it nevertheless provides
a `standard' with which we may understand the
nature of the selection effects imposed by 
resolved cluster detection, and a benchmark 
for comparing more detailed models.  
{\em It is important in the 
following that one does not forget the 
model dependence of our results, which can 
be retraced to this point of the discussion.}

\section{Unresolved Detections}

     This section is dedicated to  
the simple case of unresolved SZ detection,
which will be used as a reference in the
following discussion of 
resolved detection.  It also offers an
introduction to the main figures, Figures 
1, 2 and 3, summarizing the essential 
results of the present work.  They are
constructed for two representative cosmologies: a critical model 
$\Omo=1$, and an open model ($\lamo=0$) with $\Omo=0.3$.
For the counts and redshift distributions of Figures
2 and 3, I have used a CDM--like power spectrum with 
``shape parameter'' fixed at $\Gamma=0.25$; both
models are normalized to the present day abundance
of X--ray clusters -- $\sigma_8=0.6$ and 
$\sigma_8=1.0$ for the critical and open models,
respectively (e.g., Blanchard et al. 1999; Borgani et al. 1999;
Viana \& Liddle 1999b).

     Observations for which most clusters
are unresolved measure the total SZ flux density.
One can then simply invert
Eq. (\ref{eq:SZfluxdens}) to find the corresponding limiting 
detection mass as a function of redshift, $\MdetectUR(z,\Snu)$:
\begin{eqnarray}\label{eq:URMdet}
\nonumber
\MdetectUR(z,\Snu) & = & (0.12\times 10^{15} h^{-8/5} \msun)
   \left(\frac{\Snu}{{\rm mJy}}\right)^{3/5} \\
\nonumber
   & & (\fnu\fgas)^{-3/5}\Omo^{-1/5}\left(\frac{178}{\delNL}\right)^{1/5} \\
   & & D^{6/5}(z)(1+z)^{-3/5}
\end{eqnarray}
Integrating the mass function over redshift and over masses 
greater than this limit directly yields the source counts:
\begin{equation}\label{eq:counts}
\frac{dN}{d\Omega}(>\Snu) = \int_{0}^{\infty} dz \frac{dV}{dz d\Omega} 
        \int_{\Mdetect(z,\Snu)}^\infty dM \; \frac{dn}{dM} (M,z)
\end{equation}
The corresponding redshift distribution is simply
obtained as the integrand of the $z$--integral.

     Figure 1 compares the various detection
masses as a function of redshift for 
observations at $2$ mm, e.g., a bolometer array,
and at $1$ cm, representative of an interferometer; 
in each case the upper (red) curve corresponds to the
open model.  For the moment, concentrate
only on the the dashed lines, which 
give the result for unresolved detection, 
Eq. (\ref{eq:URMdet}), at a flux density 
of $\Snu = 1.5$ mJy.  These 
curves remain unchanged from Figure 1b to 
1c, both at $1$ cm but differing in angular 
resolution, because resolution is irrelevant for point 
sources (ignoring source confusion issues). Observe that in all
cases the detection mass {\em decreases} with
redshift beyond $z\sim 1$.  This remarkable behavior is
directly attributable to the fact that the SZ
surface brightness is independent of distance.
As already emphasized, the distance appearing in
Eq. (\ref{eq:URMdet}) is the {\em angular distance}
and not the luminosity distance,  
a factor of $(1+z)^2$ larger.  At high
$z$ the redshift dependence therefore scales
as $\sim z^{-9/5}$, one
power coming from the assumed redshift
scaling of the virial temperature and
the rest from the {\em decrease} in angular
distance (focusing) as $\sim 1/z$.
A self--similar cluster model, implicitly
assumed in this context by the constancy
of $\fgas$, thus predicts that SZ observations
are {\em more sensitive to objects at large,
rather than intermediate, redshifts}.  
This overall behavior would not change even if
we broke the self--similarity with 
a declining gas mass fraction with mass;
such a dependence could only modify the
rate of decrease with $z$.  On the other hand,
an explicit decrease in $\fgas$ with
redshift stronger than $(1+z)^{-3}$ would
cause $\MdetectUR$ to actually increase
with redshift.
It is perhaps not so surprising that at close range,
small $z$, the detection mass also drops;
this is simply due to the increasing angular
size of the object creating an increase
in total flux density (the source is assumed
to always remain unresolved in this discussion).

   From the difference between Figures 1a and 
1b,c, we see that, at a given sensitivity,
the $2$ mm observations probe farther 
down in mass.  This is nothing more than
the spectral shape of the SZ effect, described
by the function $\jnu$: the biggest decrement
occurs precisely near $2$ mm (the maximum emission
of the effect is around $750 \mu$m).  The
resolved detection mass limits, to be shortly
discussed, depend also on the angular resolution.

        Source counts for the two cosmological scenarios
are given in Figure 2.  These have been calculated using 
Eq. (\ref{eq:counts}) and the appropriate detection mass.
In order to shed some light on 
the importance of low mass objects to these results, 
the counts are presented in pairs, one curve for a 
low mass cut--off of $10^{13}\; \msun$ and one
for a cut--off of $10^{14}\; \msun$.
Note that the $x$--axis denotes the {\em pixel} noise, $\sigpix$,
and {\bf not} a limiting source flux density; in the
present situation of unresolved detections, this just
means that the corresponding limiting flux density
is $q_{\rm det}\times\sigpix$. 

        The first thing to remark from Figure 2 
is the large difference between
the two cosmological models.  The presence of clusters
at high redshift in a low--density model shows up in the
integrated counts, as confirmed by the corresponding 
redshift distributions shown in Figure 3, where the 
huge difference in cluster abundance
at large redshift is evident.  It is for
this reason that the redshift distribution of SZ sources
is a potentially powerful tool for constraining
$\Omo$ (Barbosa et al. 1996, Bartlett et al. 1998). 
This is of foremost importance and represents one of the
primary motivating factors behind this type of survey.

     This situation of unresolved 
sources applies in practice 
to missions such as the Planck Surveyor,
as discussed, for example, by
Barbosa et al. (1996) and Aghanim et al. (1997).
The higher angular resolution of possible 
ground--based surveys calls for examination
of resolved source detection.

\section{Resolved Detections}

        In this, the principle section of this {\em paper},
we treat in detail the issue of resolved SZ cluster detection.
The context will be one of arcminute resolution (pixel size) and
sub--mJy sensitivity, as targeted by the up--coming
ground--based instruments. 
It is worth being very explicit about the nature of the observations:
the simplest case to imagine corresponds to that of an
image produced by a bolometer
array, such as BOLOCAM.  In this case each point on the
image, a `pixel', represents a sample point
of the sky brightness, as transformed by the optics of the observing
system.  The optical response  may be divided into
that of the telescope--plus--atmosphere (defining the 
projection of the sky onto 
the focal plane) and the optics proper to the detector (which 
act on the focal--plane image).  There is a difference
between bolometer arrays and the familiar example  
of a CCD camera working in the visible.  For the 
latter, atmospheric seeing and telescope optics
project the sky onto the focal plane by convolving with
a Gaussian, and the camera itself then convolves this focal--plane
image with a square top--hat, one centered on each pixel. 
The difference with a bolometer array lies  
in the fact that the CCD camera
defines sharp, well--defined pixel boundaries, while a 
bolometer array, with its set of cones, convolves the
focal--plane image with something closer to a Gaussian.
This means that, unlike CCDs, the pixels of a bolometer
array `overlap' in the focal plane.  This has little
consequence for the ensuing discussion, but it is all
the same worth keeping in mind.  

         This picture is not
completely accurate when it comes to interferometers.  
Such instruments
actually directly sample the Fourier transform
of the sky.  The result may often be modeled
by a real sky image convolved with an effective,
synthesized beam, but this beam lacks sensitivity
on large scales, i.e., large spatial wavelengths
on the sky (short baselines).  Thus, the 
effective beam cannot not be precisely a Gaussian,
and it is especially important to correctly 
model the loss of response on large scales for
extended objects such as clusters.  
For the ensuing discussion, I adopt the bolometer 
picture, applying it at times rather
indiscriminately to characterize ground--based
observations; a future work will consider
the details specific to interferometric 
observations (see also the recent work of
Holder et al. 1999).  

        For a bolometer array, the response of the entire
optical chain (atmosphere-telescope-detector) 
is often adequately modeled as a bi--dimensional Gaussian
(if one is lucky, a symmetric one!), and for proper sampling,
respecting Shannon, the sample period must  be 2 -- 3 times 
smaller than the beam FWHM.  We will characterize
a survey by the pixel size and sensitivity
per pixel of its images -- $\Ompix$, a solid angle,
and $\sigpix$, a flux density.  Note that 
because the pixels `overlap' in the focal 
plane, what precisely is meant by $\Ompix$
is the square of the separation between sample
points, $\thetapix$; the concept is a bit more ambiguous than
in the case of a CCD camera.  Thus, proper sampling
means that the pixel scale $\Ompix\equiv \thetapix^2 \le \fwhm^2/4$.
It is also worth explicitly remarking that, in the 
following, I assume that the noise is {\em uncorrelated}
(from pixel to pixel) and {\em uniform} over the image.

\begin{figure}
\resizebox{\hsize}{!}{\includegraphics{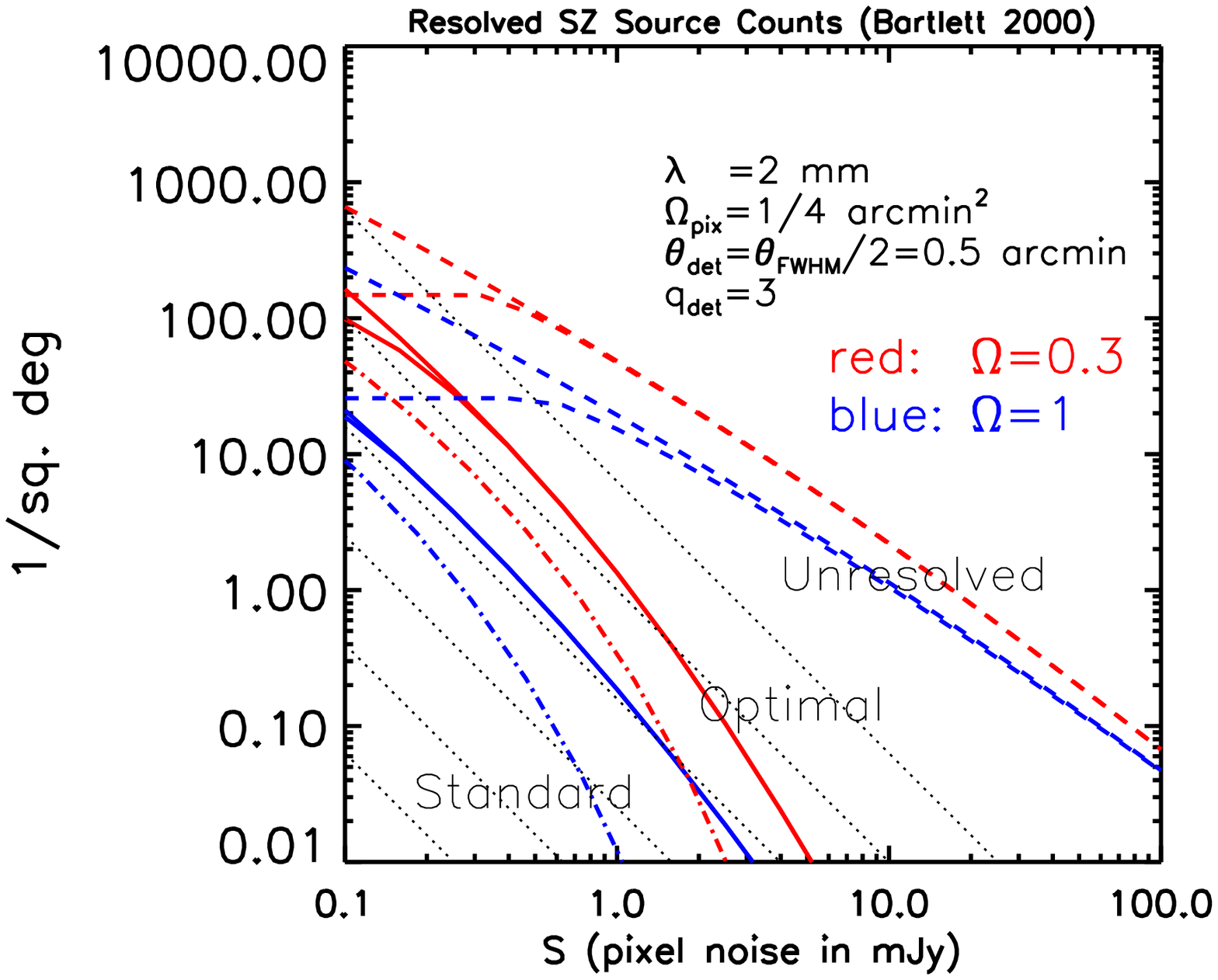}}
\caption{{\bf a)} Cluster integral source counts at $2$ mm 
as a function of map {\em pixel noise} for the two 
cosmological models introduced in the text.  The angular 
resolution and sampling correspond to the situation of Figure 1a.
Unresolved, optimal resolved and standard
resolved counts are shown, respectively, as the dashed, solid
and dot--dashed lines; the upper (red) curve in each case corresponds to
the open model with $\Omega=0.3$.  For unresolved detections,
the limiting source flux density is simply $q_{\rm det}
\times$(pixel noise).
The light dotted lines in the background indicate the critical
slope of $-2$.  The fact that the resolved counts are
steeper than this value implies that, down to low noise levels,
deep integrations yield more objects than wide and shallow ones.}
\end{figure}
\setcounter{figure}{1}
\begin{figure}
\resizebox{\hsize}{!}{\includegraphics{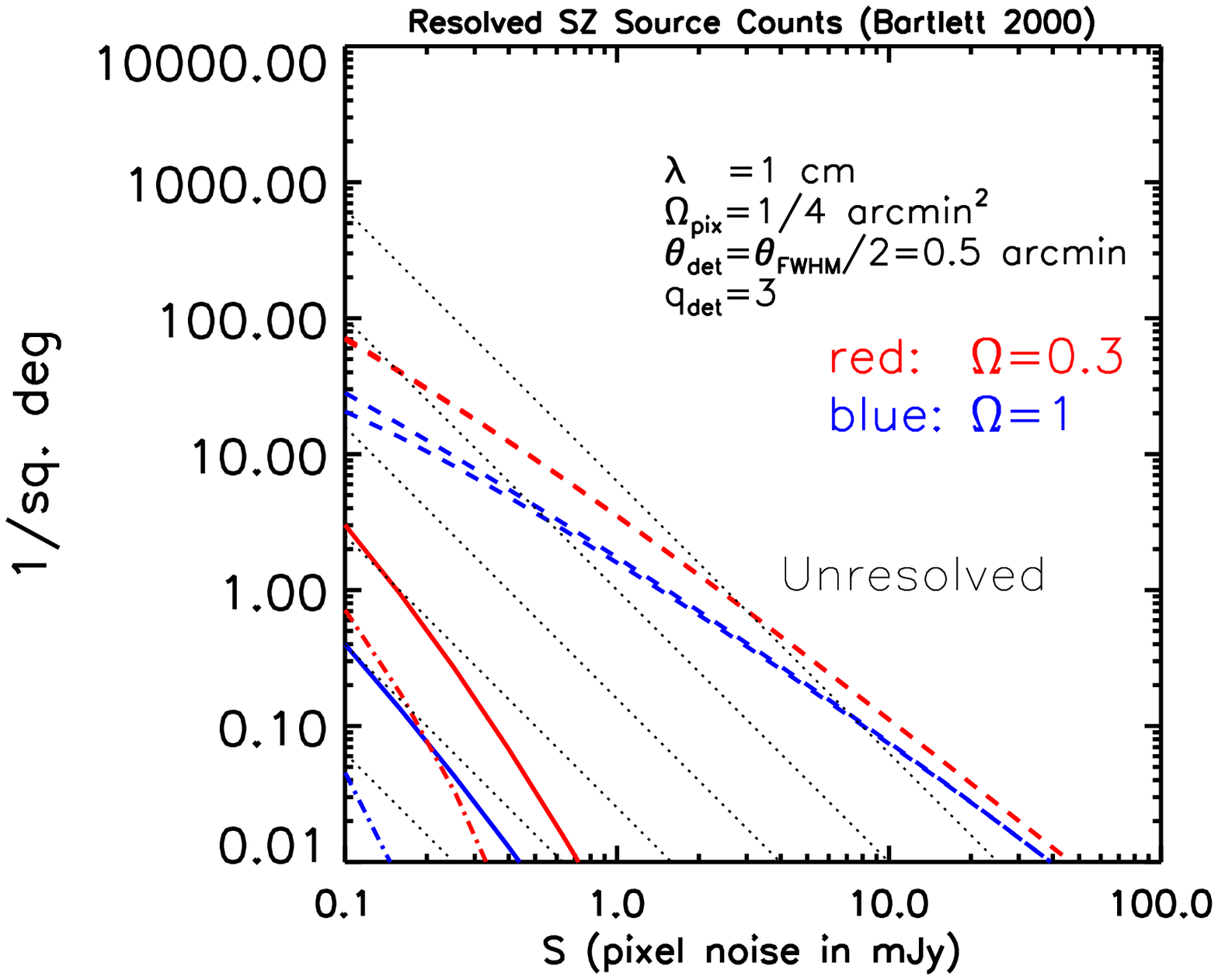}}
\caption{{\bf b)} Same as the previous figure for an observation 
wavelength of $1$ cm; curve types have the same meaning as before
(the labels `Optimal' and `Standard' have been removed for
clarity).  The resolution and sampling correspond to the
situation of Figure 1b.}
\end{figure}
\setcounter{figure}{1}
\begin{figure}
\resizebox{\hsize}{!}{\includegraphics{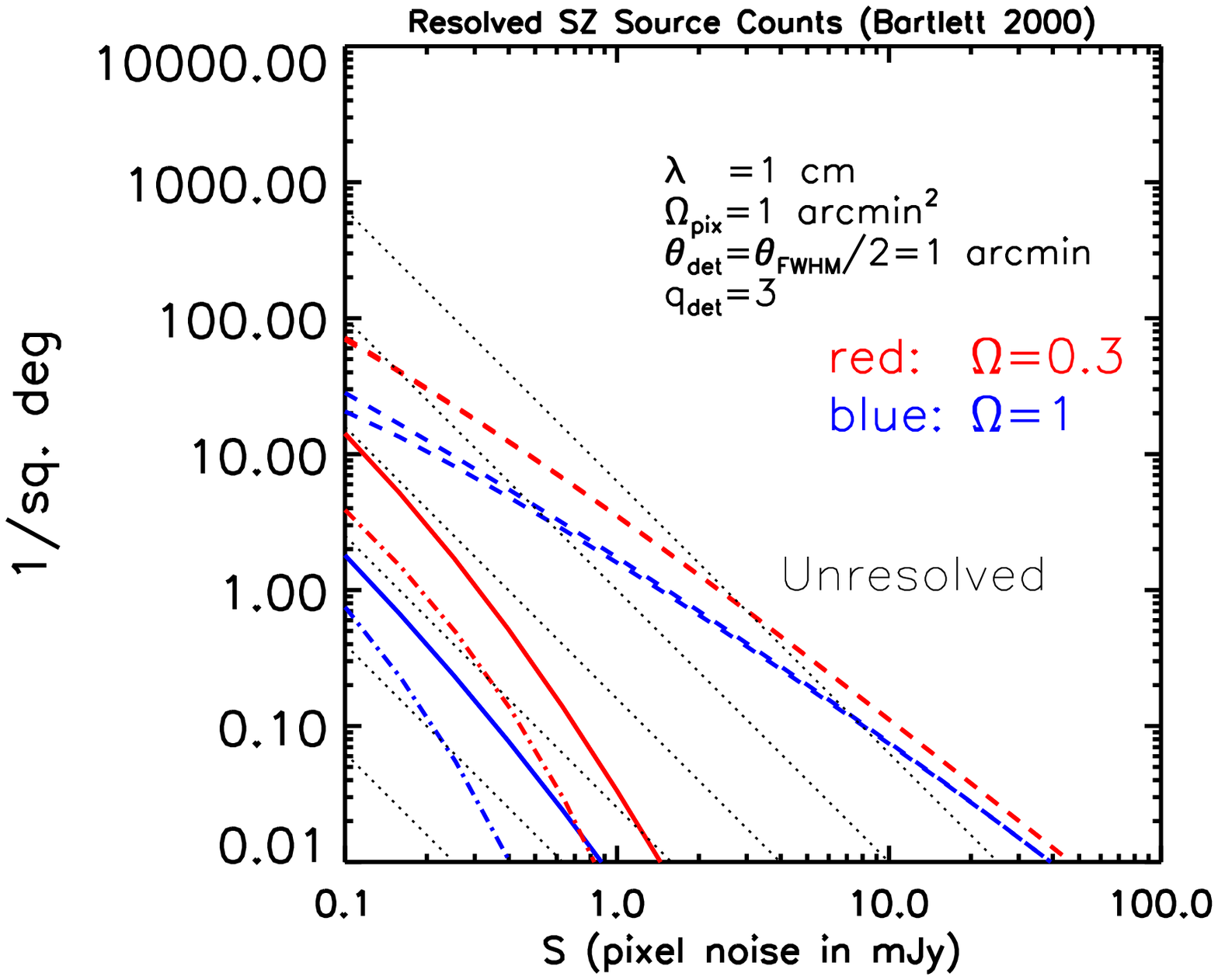}}
\caption{{\bf c)} Same as Figure 2b ($\lambda=1$ cm) but now
for $\Ompix=1$ arcmin$^2$, i.e., the situation 
of Figure 1c.  The smaller detection masses at this
lower resolution result in higher counts when 
compared to Figure 2b.  Note that the unresolved 
counts are the same here as in Figure 2b.}
\end{figure}

        Given, then, a map of the SZ sky,
we would like to understand how to extract 
clusters and the nature of the selection 
imposed by our technique.  In addition to the
observational parameters $\Ompix$ and $\sigpix$,
this will depend on the form of the extended emission 
of the sources, a complication avoided in the case
of unresolved cluster detection; this represents
an important difference between the two situations.
Employing the $\beta$--model introduced previously,
Eq. (\ref{eq:bmodel}),
we see that a cluster SZ profile may be 
described by a characteristic central
surface brightness, $\yo$, and an angular
size, $\thetac$ (the core radius).
When couched in 
terms of the purely empirical parameters
of $\Ompix$, $\sigpix$, $\yo$ and 
$\thetac$, we have before us a rather
classic and well--known problem of Astronomy.  
The only difference with galaxies in the
optical is the form of the source profile.  
All physics specific to the SZ effect itself
appears only in the relation of the 
empirical source descriptors -- $(\yo,\thetac)$ --
to the theoretically meaningful ones of cluster mass, 
$M$, and redshift, $z$.

     The procedure in the 
following is then always the same: quantify the
detection algorithm in terms of $\Ompix$
and $\sigpix$, and then translate this, via
the isothermal $\beta$--model, into a 
$\Mdetect(z;\Ompix,\sigpix)$.  I employ a 
notation where the imminently interesting independent 
variables of a function appear before the ``{\bf ;}'',
and parameterizing ones afterward.  Thus,
as written, the detection mass is primarily a function
of redshift, parameterized by the
survey properties $\Ompix$ and
$\sigpix$.  This function teaches us about the kinds of 
objects we detect, and leads directly to
the survey counts and the redshift distribution
of our clusters, via Eq. (\ref{eq:counts}).  
These latter quantities are the
key indicators of the science content of the
survey. 

        This procedure will be applied to
two source extraction methods in the following,
and the results compared to those for an unresolved
SZ survey.  
We will refer to the first as ``optimal detection'',
because it extracts sources
in such a way as to preserve the 
signal--to--noise across the entire range
of detectable surface brightness and 
source size.  This is achieved by lowering
the surface brightness limit for large
sources, possible due to the greater 
number of covered pixels.  The second
method, routinely used by such 
packages as SExtractor (Bertin \& Arnouts 1996),
searches for a minimum number of connected
pixels above a preset threshold.  The important
difference with the first technique is the
imposition of a {\em fixed} 
surface brightness limit, independent of 
source size.  The signal--to--noise
of the detections is no longer constant, 
but increases with source size.  This
technique may be considered sub--optimal in
the sense that it loses in--principle 
detectable low surface brightness sources,
a fact well appreciated in the case of 
optical galaxy surveys.

\subsection{Optimal case}

Optimal detection selects all sources with a
flux density
\begin{displaymath}
\Snu \ge \qopt N^{1/2} \sigpix
\end{displaymath}
(assuming spatially uncorrelated and uniform noise)
where $N$ is the number of pixels covered by the cluster
and $\qopt$ represents a threshold, say $\qopt\sim 3-5$;
in fact, $\qopt=S/N$, the signal--to--noise of the
detection.  Notice also that, as advertised, the
limiting surface brightness decreases with object
size: $<\inu> \sim \Snu/N\sim \qopt\sigpix/\sqrt{N}$.
One extracts in this way all objects detectable 
at a given $S/N$, and for this reason we may refer
to the method as optimal.  
The number of object pixels is simply
found as $N=\pi\thetavir^2/\Ompix$, 
where $\thetavir=\Rv/\Da$ is angular virial radius.
This permits us 
to express the detection mass as
\begin{eqnarray}\label{eq:OptMdet}
\nonumber
\MdetectOpt(z,\Snu) & = & (0.19\times 10^{15} h^{-2} \msun) 
       \left(\frac{\qopt\sigpix}{\rm mJy}\right)^{3/4} \\
\nonumber
&&       \left(\frac{{\rm arcmin}^2}{\Ompix}\right)^{3/8}
        (\fnu\fgas)^{-3/4} \Omo^{-1/2}\\
&&      \left(\frac{178}{\Delta(z)}\right)^{1/2} 
        D^{3/4}(z) (1+z)^{-3/2}
\end{eqnarray}

        As written, this criteria uses an aperture corresponding
to the full angular size of the object -- $\Snu$ is understood
to be the total SZ flux density in Eq. (\ref{eq:SZfluxdens}).
For resolved sources, one would like to chose an 
aperture which optimizes the signal--to--noise
ratio of the detection.  Interestingly, a 3D gas
profile close to $r^{-2}$, corresponding to a SZ
surface brightness $y \propto \theta^{-1}$,
results in a constant signal--to--noise with
aperture radius.  A $\beta$--model with $n\propto 
(1+r^2/\rc^2)^{-3\beta/2}$ and $\beta\sim 2/3$ 
exhibits this behavior at large radii, for example:
$y(\theta) \sim (1+\theta^2/\theta_c^2)^{-1/2}$.
In this case, the signal--to--noise of a SZ detection
increases from the center of the cluster image out
to the core radius, $\rc$, beyond which it turns over
to a constant out to the virial radius.  The
situation is different for X--ray images, where
the surface brightness 
falls off more rapidly, diving under the 
background at large radii.  From this we 
conclude that the simple criteria given above
provides in fact an optimum SZ detection (at least
as long as $\beta$ remains close to 2/3, as appears
to be the case locally).

\begin{figure}
\resizebox{\hsize}{!}{\includegraphics{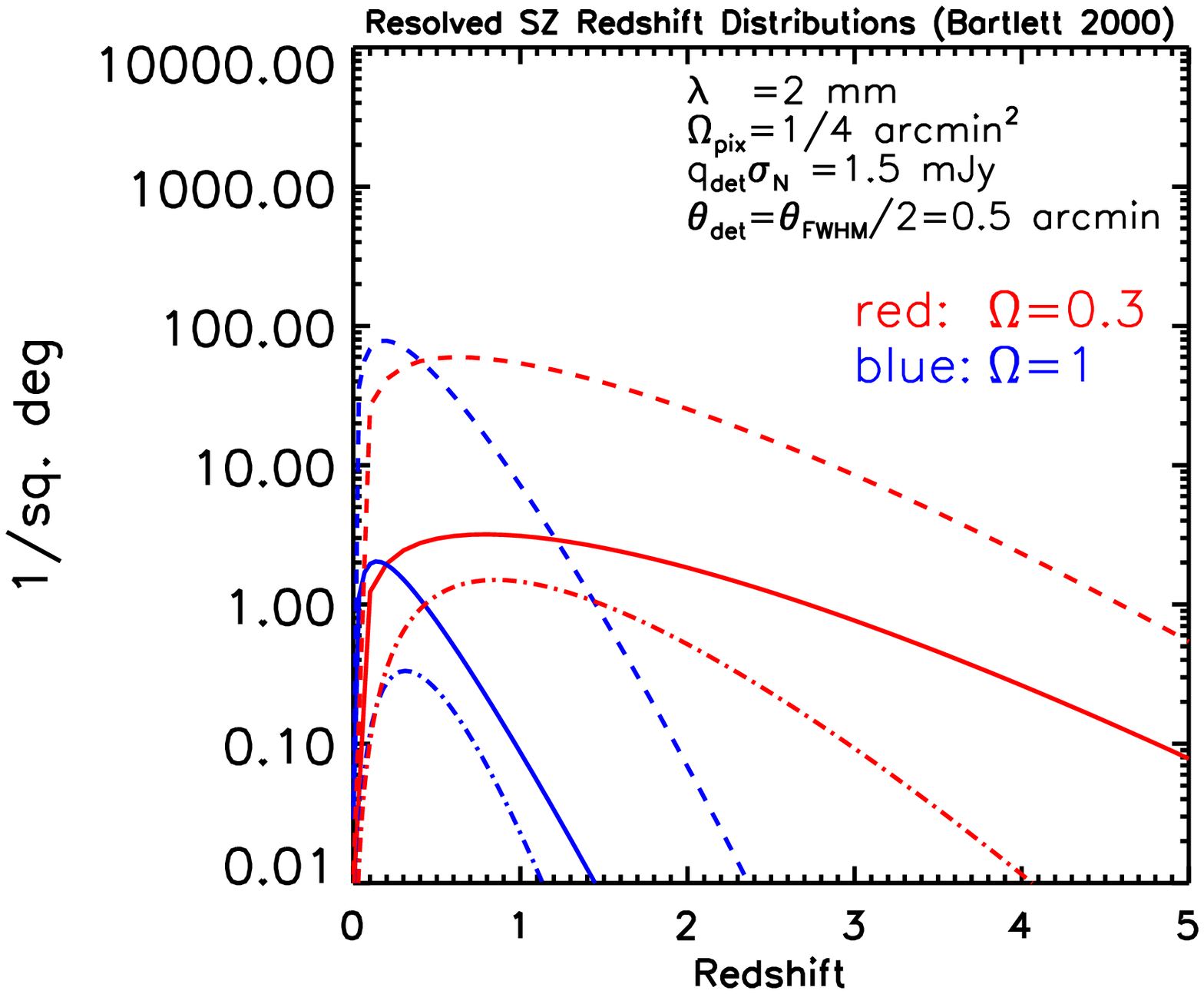}}
\caption{{\bf a)} Redshift distribution of the integrated 
counts for a flux density of $q_{\rm det}\sigpix=1.5$ mJy at $2$ mm.
The parameters are the
same as those in Figure 2a, and the line types have the
same meaning.  The upper (red) curve in each case corresponds
to the open model with $\Omo=0.3$.  Note the large difference
between the two cosmological models apparent in all cases.}
\end{figure}
\setcounter{figure}{2}
\begin{figure}
\resizebox{\hsize}{!}{\includegraphics{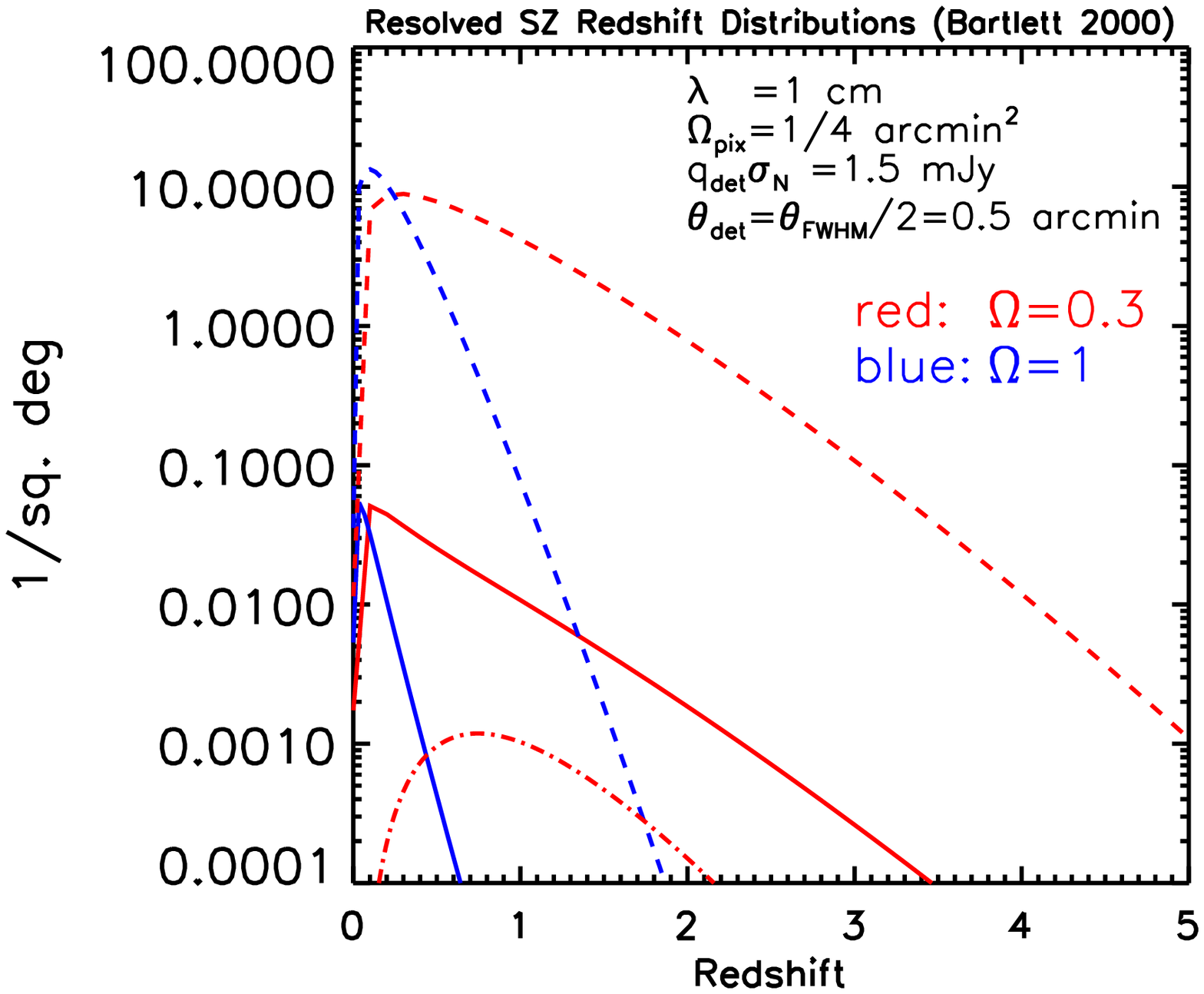}}
\caption{{\bf b)} Redshift distribution of the integrated counts
for a flux density of $q_{\rm det}\sigpix=1.5$ mJy at $1$ cm, and
for the same parameters as in Figure 2b.}
\end{figure}
\setcounter{figure}{2}
\begin{figure}
\resizebox{\hsize}{!}{\includegraphics{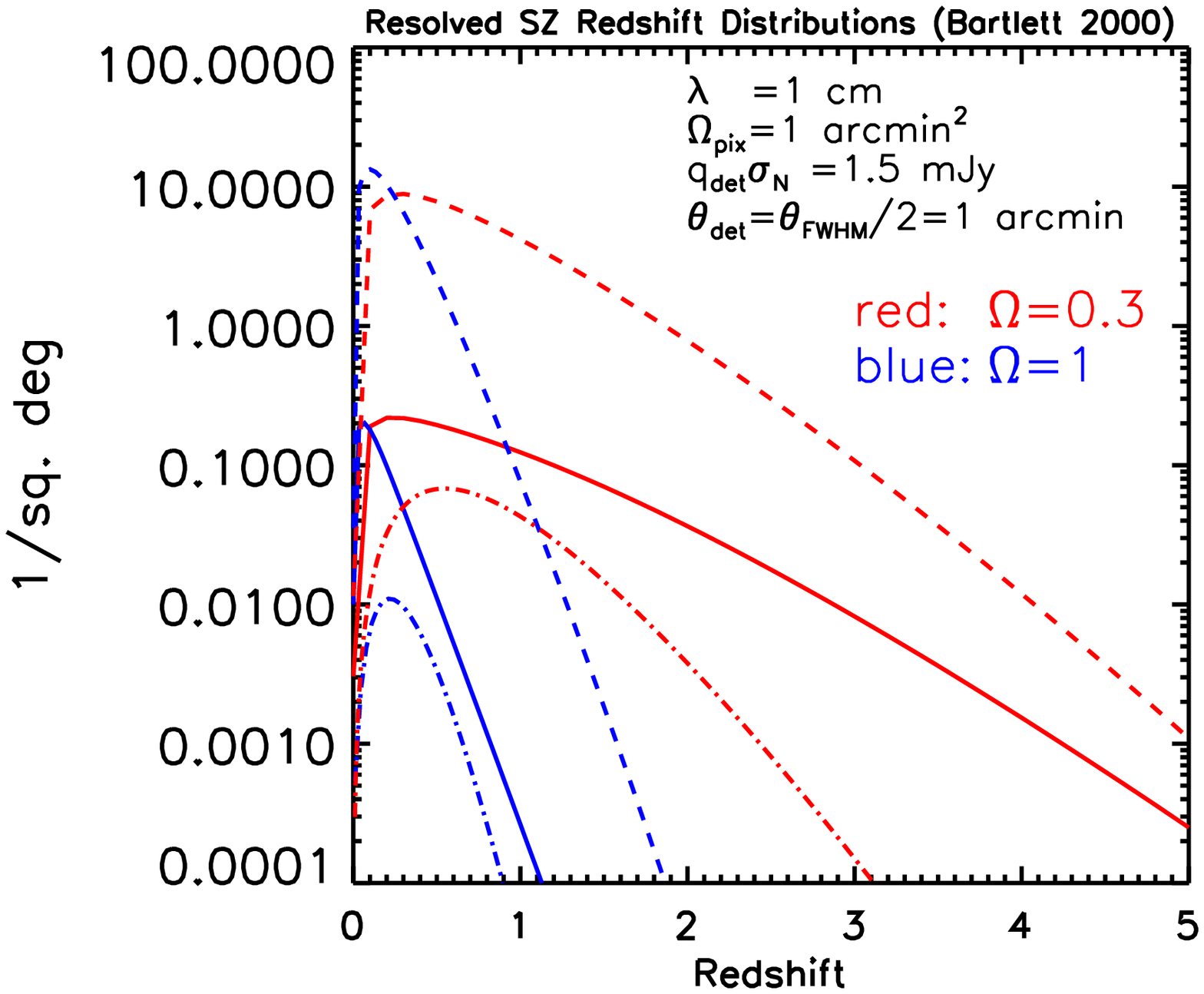}}
\caption{{\bf c)} Redshift distribution of the integrated counts
for a flux density of $q_{\rm det}\sigpix=1.5$ mJy at $1$ cm, and
for the same parameters as in Figure 2c; in particular, for
$\fwhm=2$ arcmins.}
\end{figure}

        The detection mass Eq. (\ref{eq:OptMdet}) is
displayed in Figure 1 as the solid lines.
Compared to the hypothetical point
source results, observations resolving
clusters are less efficient at detecting
clusters, especially at intermediate redshifts.
This is easy to understand as the effect
of distributing a given flux density over
$N$ pixels, {\em each} adding a noise
with variance of $\sigpix$, resulting
in a total noise level over the
object image of $\sqrt{N}\sigpix$.  A point
source, in contrast, is only subject
to the noise of one pixel, $\sigpix$.  Hence,
high resolution at fixed sensitivity 
``resolves out'' a certain fraction of
objects.  The consequences for the source
counts are clear and will be discussed 
shortly.  These curves retain the same
asymptotic behavior as before, namely
a greater sensitivity to low masses
at high redshift.  Despite the fact that
the object covers a larger number of 
noisy pixels as $z$ decreases, the optimal
method is able to take
proper advantage of the greater 
total flux density to detect low 
mass objects locally, just as in
the case of unresolved point sources.  
We shall see that this does not follow
for the standard detection routine 
(the dot--dashed lines), due to its
additional surface brightness constraint
(discussed below).  By comparing Figures 1b and 1c,
which differ only in their
angular resolution, we note that for a given 
sensitivity, lower resolution observations
are the more effective.  This is
traceable to the fact that the flux density
of a source is dispersed over fewer (noisy)
pixels than would be the case at a higher
angular resolution.  This indicates that
low resolution observations at a given
wavelength and sensitivity are to be
prefered, at least for detection purposes.
There is, however, a limit set by eventual
source confusion.

        We have just seen from Figure 1 that low 
surface brightness clusters are ``resolved out'' at high 
resolution.  This leads to overall lower counts that
are also much {\em steeper} than the equivalent for 
unresolved point sources.  
Generally speaking, the unresolved counts do not deviate
too much from a Euclidean law, $\propto \Snu^{-3/2}$; on
the other hand, the resolved counts can be much steeper.
The examples shown in Figure 2 are in fact steeper 
than $\Snu^{-2}$, indicated by the
dotted lines, down to essentially the faintest flux levels
attainable in immediately foreseeable 
observations.  This is critical for optimizing
an observing strategy with a fixed amount telescope
time, $T$.  Consider the common situation in which
the final map noise  decreases with
integration time as $1/\sqrt{t}$; then, 
the solid angle covered in time $T$, with individual
field integrations of duration $t$, scales 
with sensitivity as $ \sim  T/t \sim \sigpix^2$.
Hence, if the integrated source counts are
steeper than $\sigpix^{-2}$, one gains
objects by ``going deep'', integrating longer
on each individual field, rather than 
``going wide'', with shorter integrations
covering a larger total solid angle. 
The important conclusion to draw from Figure 2 is 
then that the way to optimize the number
of detected objects in a survey with arcminute resolution 
is by ``going deep'', down to the point were
the counts begin to flatten out.  In our examples,
this does not occur until the very lowest
flux levels deemed at present reasonable.  
It should be emphasized that this
conclusion rests on the results calculated here 
in the context of a self--similar cluster population.
It is all the same suggestive and important 
in the fact that it is contrary to the 
conclusion one would draw based on unresolved source
count calculations. 
The opposite holds
for surveys with low angular resolution where the majority
of sources remain unresolved, such as the Planck Surveyor
observations.  

     Finally, as to be expected from the ``loss'' of 
objects at intermediate redshifts, the redshift
distribution for optimally selected objects lies
under the corresponding point source examples, and
is somewhat flatter.  All the same, the two cosmological
models are easily distinguished with an enormous
``leverage'' at high $z$.

%        The corresponding redshift distributions of the integral
%counts above 1 mJy pixel noise are shown in Figure 7.  Due
%to the higher detection masses at low redshifts, where
%objects are more likely to be resolved, these distributions
%in redshift tend to be flatter than their unresolved counterparts.
%The huge difference between cosmologies at these flux
%levels is ever-present.
%True SZ surveys will be one of the best ways to search for
%virialized objects at large redshift, whose existence depends
%critically on the density parameter, as noted above.  This
%is of foremost importance and represents to our mind the
%primary motivating factor behind these surveys.
%While is true that the existence of SZ detected objects
%at high $z$ relies on the presence of free gas heated to
%the virial temperature of the collapsed halos, so this
%might be considered somewhat model dependent. For these
%calculations, we have assumed constant gas mass fractions.
%However, the differences at these flux densities are so
%large that it would require extremely rapid evolution
%of the gas mass with mass and/or redshift to eliminate it.
%As the gas is thought to be essentially primordial in 
%origin, from infall during collapse, there is no
%apparent reason for such rapid evolution.  In any case,
%the argument always works in the following direction.
%The absence of objects at large $z$ could be tolerated
%in a low-density model by eliminating the gas, however,
%the presence of such objects at high $z$ is virtually
%impossible to explain in any way in a critical model --
%the halos just don't exist!

\subsection{Standard algorithms}

        Standard detection routines typically identify sources
as a minimum number of contiguous pixels all above a preset
threshold, usually $\qst$ times the pixel noise $\sigpix$.
This is not the same criteria as above, in the optimal 
case, because we have now established a {\em fixed} surface brightness 
threshold -- $\qst\sigpix/\Ompix$ -- independent of object size 
(or luminosity).  Previously, 
we allowed ourselves to lower this threshold for larger sources,
in order to pick--up low surface brightness objects 
while maintaining a constant signal--to--noise;
for this reason, it was an optimum detection algorithm.
Here, the surface brightness is instead
a fixed, while the signal--to--noise 
increases with object size as 
$S/N = \qst\sqrt{N}$.  A further difference is that
the surface brightness cut imposes a minimum detectable 
mass at $z=0$.
We obviously expect this method to detect fewer 
objects than the optimal approach.

        Consider application of the standard algorithm
to a SZ profile, empirically described 
in the $\beta$--model by $\yo$ and $\thetac$.
Although our final goal is to understand the
selection on mass and redshift imposed by the detection 
criteria, it is quite useful, firstly, to gain
insight into the workings of detection in terms
of $\yo$ and $\thetac$.  As mentioned, what is actually recorded
at each sample point
(pixel), say by a bolometer camera,
is the sky signal integrated over the beam $B$,
which we will take to be axially symmetric.  Thus, for a pixel
at position $\hn$ (a unit vector on the sphere):
\begin{displaymath}
\Sobs(\hn) = \int d\Omega' \inu(\hn') B(\hn\cdot\hn')
\end{displaymath}
For our calculations, we shall furthermore adopt a Gaussian beam, so
that a cluster appears as a $\beta$--profile smeared by
a Gaussian of dispersion $\sigb = \fwhm/\sqrt{8\ln 2}$,
\begin{displaymath}
B = e^{-\theta^2/2\sigb^2} %\frac{1}{2\pi \sigb^2} e^{-x^2/2\sigb^2}
\end{displaymath}
where $\theta$ is the angle from the beam axis and 
$\fwhm$ is the beam full--width at half--maximum.  
Notice that we take the image to be in flux density units.
By placing the coordinate origin at the cluster center, so that
now $\hn$ is simply marked by the 
angular distance $\vec{\theta}$ from the origin
(small angle approximation),
the beam--smeared profile of a cluster may be written
as
\begin{displaymath}
\Sobs(\theta) = \yo\thetac^2 \jnu {\calG}[\theta/\thetac;\sigb/\thetac]
\end{displaymath}
explicitly separating out a dimensionless function
\begin{eqnarray*}
{\calG}(r;p) & \equiv & \int_0^\infty dx\; 
        x e^{-\frac{1}{2}\left(\frac{x}{p}\right)^2} \\
&&        \int_0^{2\pi} d\phi\; \frac{\Theta[\xv^2-r^2-x^2-2xr\cos\phi]}
        {\left(1 + r^2+x^2+2xr\cos\phi\right)^\alpha}
\end{eqnarray*} 
parameterized only by the ratio $p=\sigb/\thetac$.  The 
Heavyside function, $\Theta$, cuts off the integral
beyond the virial radius.

      It is this smeared profile of a cluster that is subject to 
the detection criteria that a minimum number of connected 
pixels, $\Nmin$, must lie
above the threshold $\qst\sigpix$.  This amounts to demanding that
the object image above a flux density of $\qst\sigpix$ cover
a minimum solid angle of $\Nmin\Ompix$.  Let $\thetadet$ be the
angular size of a cluster above the detection threshold, which
may be calculated as the root of the following equation:
\begin{equation}\label{eq:thetadet_root}
\Sobs(\thetadet) = \qst\sigpix
\end{equation}
We will say that a cluster is detected if $\thetadet$ is
large enough to cover $\Nmin$ pixels.  In the present 
analytic treatment, we will simply impose a lower limit
to $\thetadet$ and ignore any complications arising from
the discreteness of the image.  
%\begin{figure*}
%\begin{center}
%\resizebox{\hsize}{!}{\includegraphics[totalheight=17cm,width=10cm,angle=90]
%{theta_detyothetac_try.ps}}
%\end{center}
%\caption{The behavior of $\thetadet/\thetac$ as a function of
%$\thetac/\sigb$ and $\yo$:, a) for $|\tildy|=9.48\times 10^{-5}$, e.g.,
%$\lambda=2$ mm, $\fwhm=1$ arcmin and $q\sigpix=1.5$ mJy, and b)
%for $|\tildy|=6.84\times 10^{-4}$, e.g., 
%$\lambda=2$ mm, $\fwhm=1$ arcmin and $q\sigpix=11$ mJy, or
%$\lambda=2$ mm, $\fwhm=2.69$ arcmins and $q\sigpix=1.5$ mJy
%$\lambda=1$ cm, $\fwhm=1$ arcmin and $q\sigpix=1.5$ mJy
%}
%\end{figure*}
Using the function $\calG$, Eq. (\ref{eq:thetadet_root})  
may be written in compact form as
\begin{equation}\label{eq:thetadet_dimless}
\calG[\thetadet/\thetac;\sigb/\thetac] = \frac{\qst\sigpix}{\yo\thetac^2\jnu} 
                          \equiv \frac{\tildy}{\yo}
                                \left(\frac{\sigb}{\thetac}\right)^{2}
%                         \equiv \calR
\end{equation}
introducing the parameter $\tildy\equiv (\qst\sigpix)/(\sigb^2\jnu)$
characterizing the experimental set--up.
It is clear that the solution will be given as 
$\thetadet/\thetac$, that it will be a function of
$\sigb/\thetac$ and $\yo$, and that it will be
parameterized by $\tildy$, i.e.,
\begin{displaymath}
[\thetadet/\thetac](\sigb/\thetac,\yo; \tildy)
\end{displaymath}
To understand the role of $\tildy$, study  
the result in the limit as $\thetac\rightarrow \infty$;
this will be particularly important below,
when we consider the non--zero detection mass
at zero redshift imposed by the surface 
brightness cut.  In this large--object 
limit, $\calG(r;p) \rightarrow \calG(r\rightarrow 0;p\rightarrow 0)
\rightarrow 2\pi p^2$.  We thus find that
in order to be detected an object must have
a central surface brightness
\begin{equation}\label{eq:ylimit}
\yo > \frac{\tildy}{2\pi}
\end{equation}
In other words, $\tildy$ indeed embodies the 
surface brightness cut.  This will be used
shortly.

\begin{figure}\label{}
\resizebox{\hsize}{!}{\includegraphics{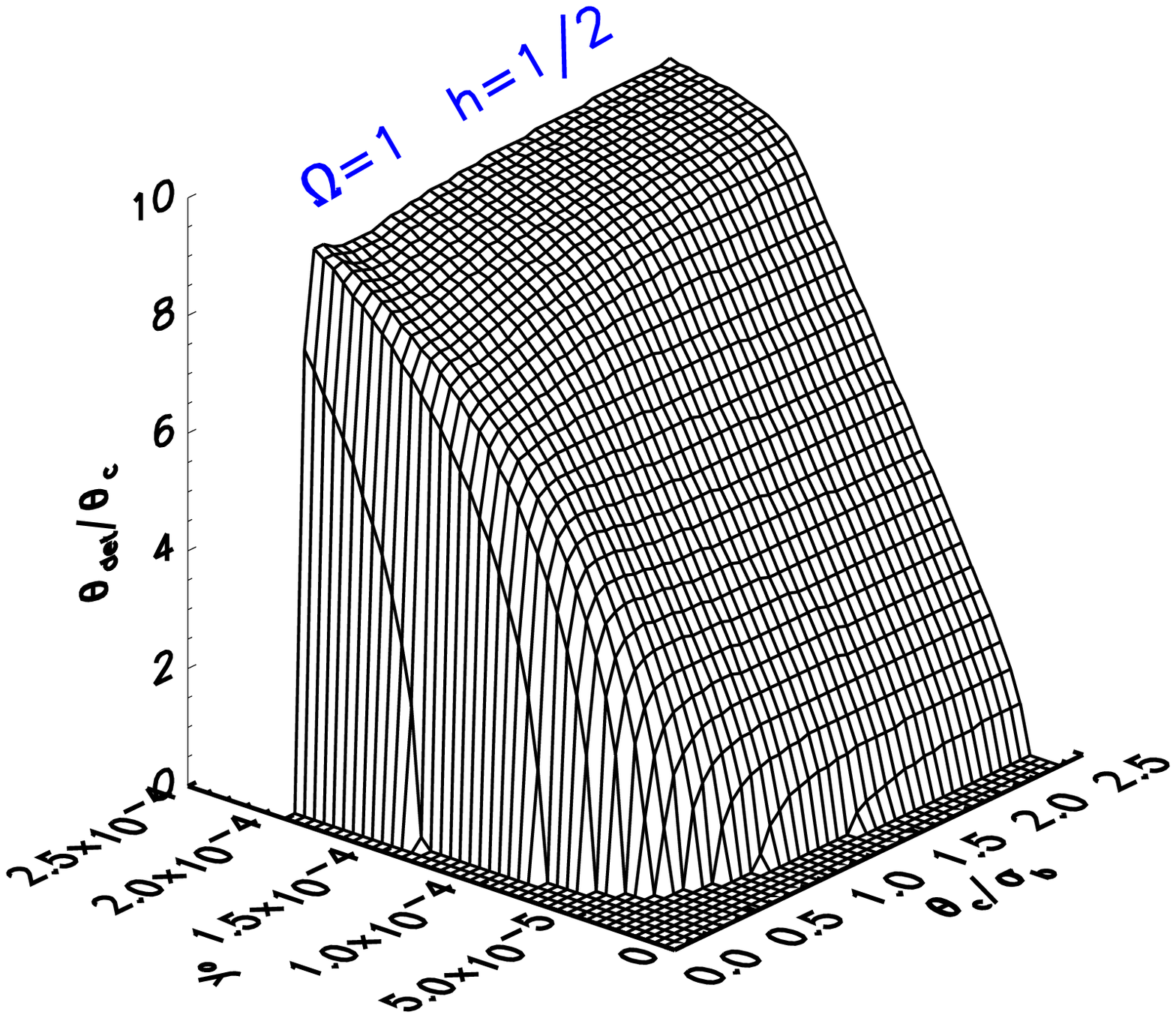}}
\caption{{\bf a)} The behavior of $\thetadet/\thetac$ as a function of
$\thetac/\sigb$ and $\yo$ in a critical--density universe ($\Omega=1$,
$h=1/2$) and for $|\tildy|=9.48\times 10^{-5}$, e.g.,
$\lambda=2$ mm, $\fwhm=1$ arcmin and $q_{\rm det}\sigpix=1.5$ mJy.}
\end{figure}
\setcounter{figure}{3}
\begin{figure}\label{}
\resizebox{\hsize}{!}{\includegraphics{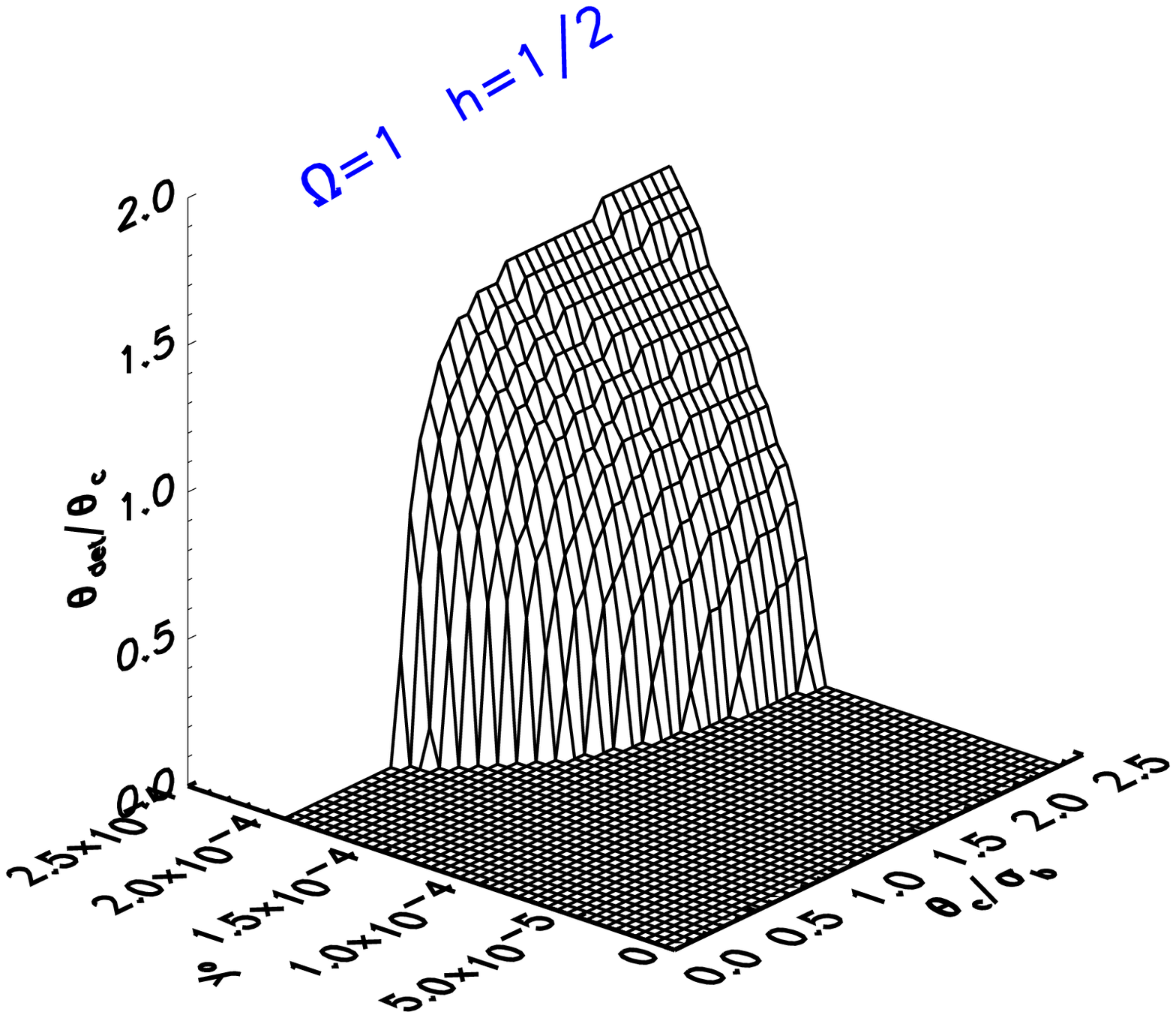}}
\caption{{\bf b)} Same as Fig. 4a, but for 
$|\tildy|=6.84\times 10^{-4}$, e.g., 
$\lambda=1$~cm, $\fwhm=1$~arcmin and $q_{\rm det}\sigpix=1.5$~mJy
(or 
$\lambda=2$~mm, $\fwhm=1$~arcmin and $q_{\rm det}\sigpix=11$~mJy, or
$\lambda=2$~mm, $\fwhm=2.69$~arcmins and $q_{\rm det}\sigpix=1.5$~mJy).
}
\end{figure}

%       We, however, are intent on understanding the
%detection criteria in terms of our empirical variables
%$\yo$ and $\thetac$, for which it is instructive to 
%express $\calR$ in terms of a parameter 
%$\tildy\equiv (q\sigpix)/(\sigb^2\jnu)$ 
%characterizing the experimental set--up:
%\begin{displaymath}
%{\calR} = \frac{\tildy}{\yo}\frac{\sigb}{\thetac}
%\end{displaymath}
%Now the 
%solution should be considered as a function of $\sigb/\thetac$ and 
%$\yo$ and as parameterized by $\tildy$: 
%$[\thetadet/\thetac][\sigb/\thetac,\yo;\tildy]$.
%$|\tildy|=9.48\times 10^{-5}$, e.g., $\lambda=2$ mm, $\fwhm=1$ arcmin,
%$\sigpix=0.5$ mJy and $q=3$, and for $|\tildy|=6.48\times 10^{-4}$, e.g.,
%$\lambda=1$ cm, $\fwhm=1$ arcmin,
%$\sigpix=0.5$ mJy and $q=3$. 

     Figure 4 shows the solution over the 
$(\thetac/\sigb,\yo)$--plane for two reasonable values of 
$|\tildy|$. To understand this figure, separate the plane into a  
region occupied by resolved sources -- $\thetac/\sigb >>1$ -- 
and the region of point sources,  $\thetac/\sigb << 1$:
\begin{itemize}
\item {\em Resolved sources}:  It is clear that by
increasing $\yo$ at fixed $\thetac/\sigb (>>1)$,
we see an ever increasing portion of the ICM.  
The cluster `lights-up' 
until we see all of it, out to the virial radius (beyond
which we assume that the gas has not been heated), and
the solution flattens out at this point to the 
adopted value $\xv=\Rv/\rc=10$; beyond this,
there is no more cluster to be seen.  
Of course, in the other direction, the object is
eventually lost as we decrease $\yo$ to the point where
even the central parts of the cluster do not rise above
the detection threshold.  
%At fixed $\yo$, high enough that the cluster is detectable, 
%we see that $\thetadet\sim \xv\thetac$ (always within the 
%resolved region), which makes sense: 
%the size of the detected region simply scales with the 
%cluster size, when it is resolved.  

\item {\em Point--source limit} ($\thetac/\sigb \rightarrow 0$):
In this extreme, the source profile becomes that of the beam, 
normalized to the total source flux density.  This latter quantity
scales as $\yo\jnu\thetac^2$, so that as $\thetac/\sigb$ continues 
to decrease at fixed $\tildy$ (i.e., holding $\sigb$ constant), 
the imprint of the object gradually sinks below the detection 
threshold and $\thetadet\rightarrow 0$; this explains
the cut-off at low $\thetac/\sigb$ for a given 
central surface brightness.
\end{itemize}

\begin{figure}
\resizebox{\hsize}{!}{\includegraphics{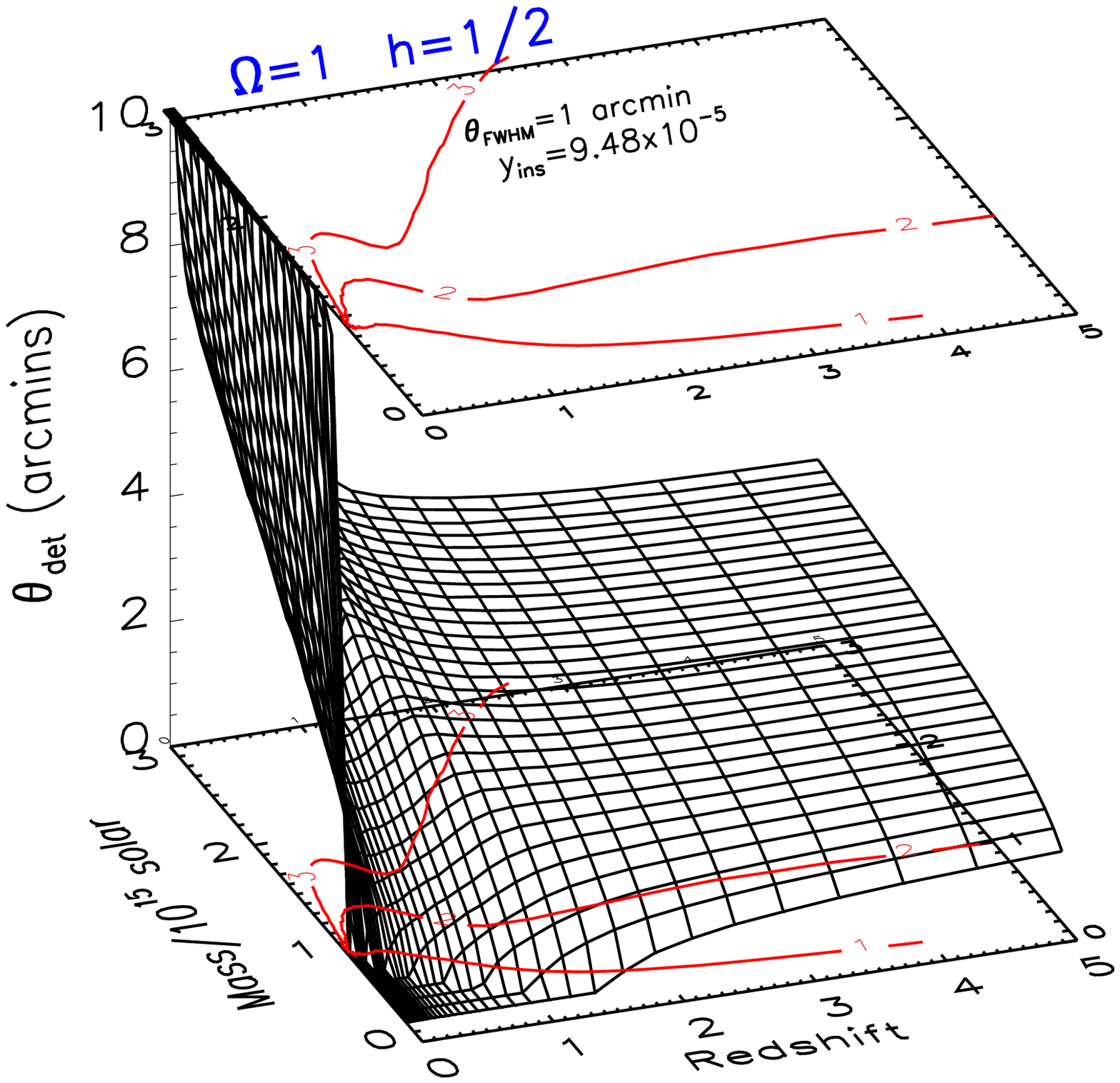}}
\caption{{\bf a)} Source detection radius (arcmins) as a function of cluster
mass $M$ (units of $10^{15}$ solar masses) and redshift, for
$\fwhm=1$ arcmin and $\tildy=9.48\times 10^{-5}$,
e.g., $\lambda=2$ mm and $\qst\sigpix=1.5$ mJy.  Notice the
non--zero detection mass at $z=0$.}
\end{figure}                    
\setcounter{figure}{4}
\begin{figure}
\resizebox{\hsize}{!}{\includegraphics{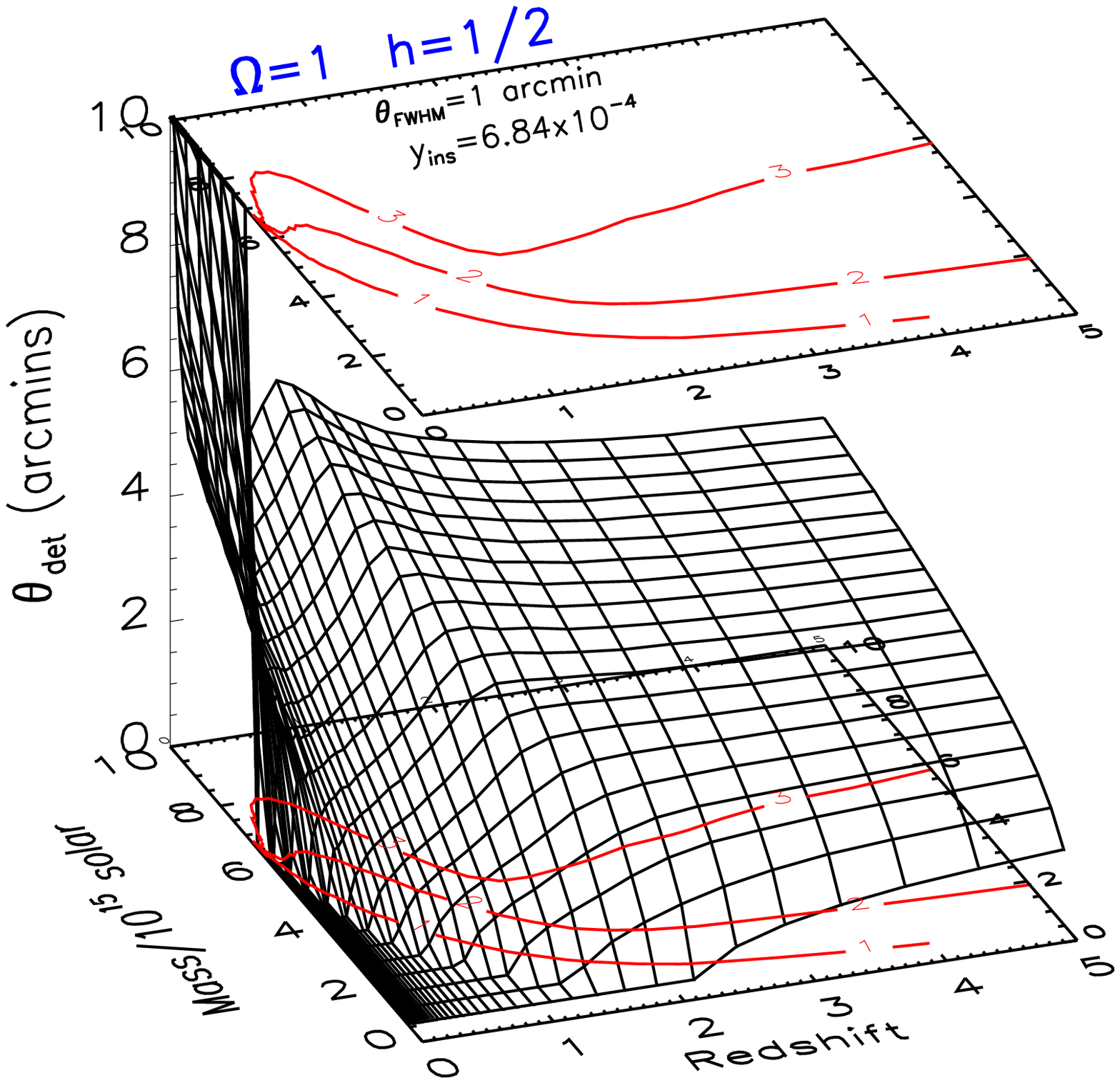}}
\caption{{\bf b)} Source detection radius for
$\fwhm=1$ arcmin and $\tildy=6.84\times 10^{-4}$,
e.g., $\lambda=1$ cm and $\qst\sigpix=1.5$ mJy.  Note the
change of scale for the $M$--axis relative to the previous figure.}
\end{figure}                    
\setcounter{figure}{4}
\begin{figure}
\resizebox{\hsize}{!}{\includegraphics{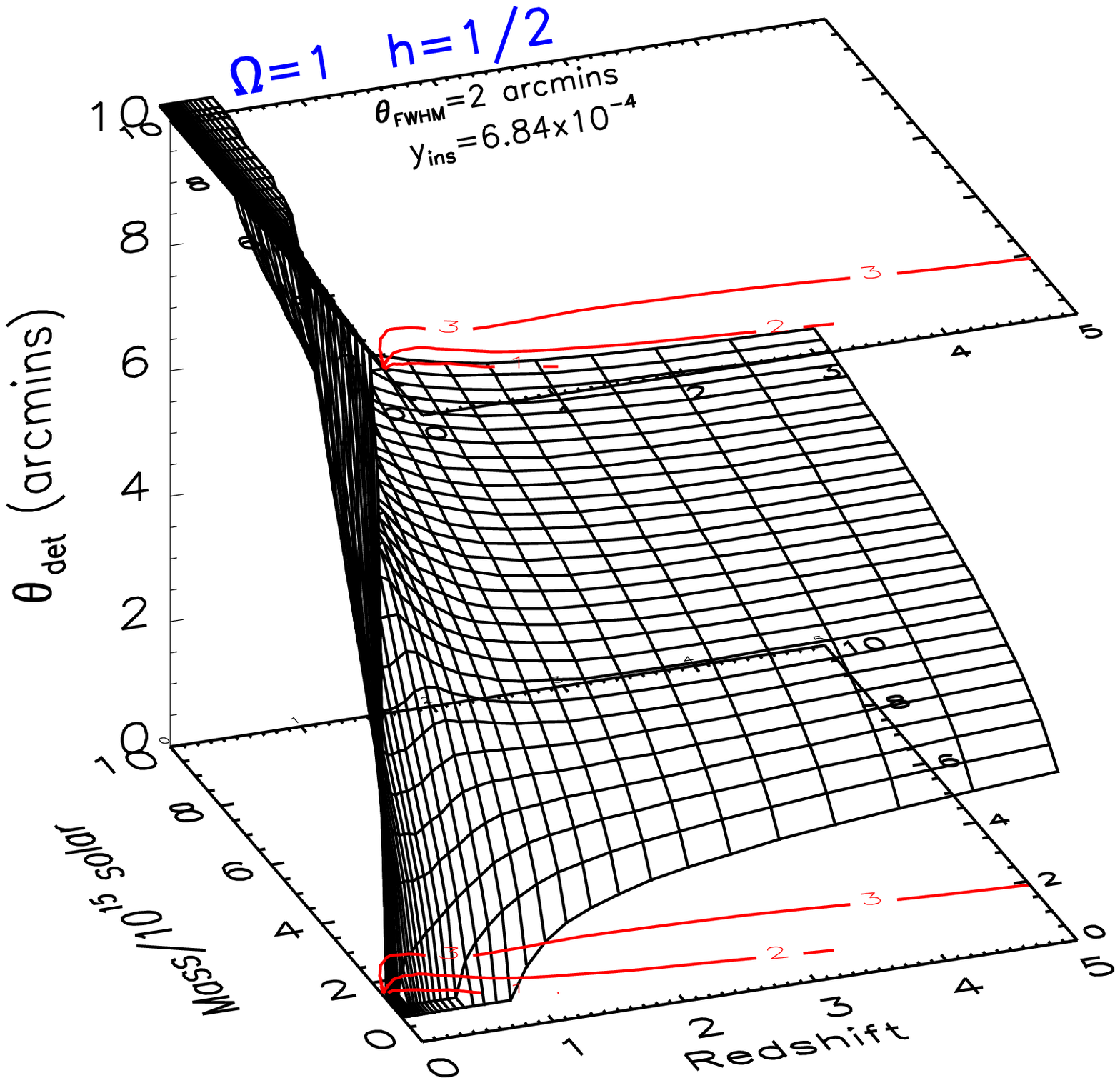}}
\caption{{\bf c)} Source detection radius for
$\fwhm=2$ arcmins and $\tildy=6.84\times 10^{-4}$,
e.g., $\lambda=1$ cm and $\qst\sigpix=1.5$ mJy.}
\end{figure}                    

        So far, nothing extraordinary, but rather the standard
issues of extended object detection given a particular intensity
profile.  Modeling more specific to the SZ effect enters only when we
apply the SZ--based relations between the central surface brightness 
and angular size -- $\yo$ and $\thetac$ -- and the 
theoretically meaningful quantities of cluster mass, $M$,
and redshift, $z$.  These relations allow us to translate
a surface like that of Figure 4 into an equivalent surface 
over the $(z,M)$--plane, as shown for three different cases 
in Figure 5 (all using our adopted self--similar cluster model).
%with $(\lambda,\fwhm,q,\sigpix) = (0.2{\rm\ cm},
%1{\rm\ arcmin},3,0.5{\rm\ mJy})$.  
Note that these latter surfaces (Figure 5) are not uniquely
parameterized by $\tildy$,
because the translation from the axes
in Figure 4 to mass and redshift explicitly 
involves $\sigb$.  Thus, there are now two governing parameters, which
we can take to be $\tildy$ and $\sigb$:
\begin{displaymath}
\thetadet[z,M;\tildy,\sigb]
\end{displaymath}
%and $\tildtheta\equiv
%\sqrt{\tildy\sigb^2} = \sqrt{(q\sigpix/\jnu)}$,
%a characteristic angle: $\thetadet[z,M;\sigb,\tildtheta]$.

        To study Figure 5 in detail, recall the simple scaling 
relations $\yo \sim nM \sim M\delNL(z) (1+z)^3$ and $\thetac 
\sim M^{1/3}\delNL^{-1/3}(z)(1+z)^{-1}/\Da(z)$, valid
if the core radius scales with virial radius
(it may not, but this has been assumed in the construction of 
the figure).  Notice that mass and redshift are mixed in a nontrivial
way in the expressions for surface brightness and angular extent.
In particular, a cluster of given mass becomes more centrally
bright towards higher redshift, due to a higher gas density (scaling
$\sim$ the background), while its angular extent at first 
decreases rapidly, as $\sim 1/z$ at low redshift, and 
then approaches an asymptote, since $(1+z)\Da(z)\rightarrow 
2c(\Ho\Omo)^{-1}$ towards large $z$.

      Consider firstly the low--redshift region of Figure 5b
(the various effects are most clearly displayed in panel b),
where the $z$-dependence is dominated by $\Da$; here, 
mass uniquely parameterizes the surface brightness, i.e., $\yo$,
while $z$ changes only $\thetac$: 
\begin{itemize}
\item At constant $M$, 
decreasing $z$ increases the angular size of a cluster,
so that, for resolved objects that are not completely below the 
surface brightness detection threshold ($\qst\sigpix/\Ompix$), 
$\thetadet$ rises as $1/z$; this corresponds to the 
$\thetadet/\thetac\propto const$ behavior in Figure 4.  
For smaller objects, of low mass, the beam profile determines
the scaling of $\thetadet$ with $z$, and this corresponds to 
the point--source limit of Figure 4.

\item At fixed (low) redshift, low mass
objects eventually fall below the surface brightness limit, and
$\thetadet$ reaches zero; on the other hand, massive
clusters are `lit--up' out to their virial radius, at which point
$\thetadet$ attains $\thetavir=\xv\thetac$, which grows as $M^{1/3}$. 
\end{itemize}
An important aspect of detection procedure, mentioned
at the beginning of this section and now evident from 
Figure 5, is the existence of 
a minimum detectable mass in the limit of zero redshift
(on the $M$--axis).  This characteristic mass is established 
by the surface brightness threshold -- $\qst\sigpix/\Ompix$,
and its existence represents
a fundamental difference with the previous case of optimal
detection, where arbitrarily low mass (low surface brightness)
clusters where picked up if they were large enough, i.e., 
very close at $z=0$.  We already saw in Eq. (\ref{eq:ylimit}) 
how $\tildy$ summarizes the surface brightness constraint.
The low redshift detection mass limits seen in Figure 1 and
here in Figure 5 are indeed reproduced numerically 
from Eq. (\ref{eq:ylimit}), once Eq. (\ref{eq:yo}) is used to 
convert $\yo$ into a mass at $z=0$.  

        At redshifts approaching unity and
beyond, $M$ and $z$ are fully mixed in the expressions for 
$\yo$ and $\thetac$:  
\begin{itemize} 
\item Massive clusters well above
the surface brightness threshold
increase in surface brightness with redshift 
to the point where they are completely seen, all the
way out to $\thetavir$; at even larger $z$, $\thetadet$
reflects the gradual fall--off to the 
asymptote set by the angular--size distance. This 
explains the ridge running down the surface 
in Figure 5 around $z=1$. Less massive 
clusters, on the other hand, only reach 
the point of full illumination at higher
$z$, well into the asymptotic behavior of
$\thetavir$, and hence the ridge tends to
be washed out at the low mass end.  Finally, the central 
surface brightness of very low mass clusters 
falls below the detection threshold at 
ever larger redshifts, i.e., the boundary $\thetadet=0$
moves outward in $z$ as $M$ is decreased.
\end{itemize}

     Compare now the three panels of Figure 5.
We observe the greater sensitivity at $2$ mm,
due to the spectrum of the SZ effect, by
the fact that a given cluster of mass $M$ and 
$z$ produces a smaller $\thetadet$ at $1$ cm 
wavelength in panel b (notice the change in scale along
the $M$--axis between panels a and b).
The same remarque applies to the greater 
sensitivity, at a given noise level, of
the lower resolution observations exemplified
in panel c.  These characteristics will 
be inherited by the detection mass curves,
our next topic.

\subsubsection{Detection mass as a function of redshift}

        Since $\thetadet$ increases monotonically with
$M$, the contours displayed on the top and bottom faces of 
Figure 5 represent curves of minimum detectable 
mass, $\MdetectSt(z;\thetadet)$, each one for a {\em different 
detection threshold} defined by {\em different values of} 
$\thetadet$ (indicated in arcminutes on the contours 
in the figure).  All of these contours, however,
are defined for the same value of $\qst\sigpix$ set
by the governing parameters $\tildy$ and $\sigb$ (or $\fwhm$).
In contrast to the optimal routine, a detection in
the standard case is specified by not one, but
two parameters -- the pair $(\qst,\thetadet)$.  
This embodies the fact that a detection must
satisfy two criteria: a minimum flux density,
$\sim \qst\sigpix\thetadet^2/\Ompix$, {\bf and} a 
minimum surface brightness, $\sim \qst\sigpix/\Ompix$.
In practice, the choice of values for $\qst$ and
$\thetadet$ may be somewhat of
a black art, but once made it specifies
the survey's characteristic $\MdetectSt(z)$. 
For the ensuing examples, I make the
choice motivated by the following considerations:
Note that the signal--to--noise
of a detection $S/N = \qst\sqrt{\Nmin}$.  The
detection criteria imposed as  
$\thetadet = \sqrt{\Nmin}\thetapix$ may
then be expressed in terms of the $S/N$:
\begin{equation}\label{eq:thetadet}
\thetadet = \frac{1}{\qst}\left(\frac{S}{N} \right)
            \left(\frac{\thetapix}{\fwhm} \right) \fwhm
\end{equation}
For reference, recall that in the optimal approach the
parameter $\qopt$ was exactly the $S/N$.
Now, at fixed signal--to--noise, a larger $\qst$ leads to a smaller
$\Nmin$ (i.e., $\thetadet$), which facilitates the detection
of fainter point sources, because their flux density is
buried in less noise (fewer pixels).
On the other hand,
a large value of $\qst$ disfavors finding low surface
brightness objects; thus, a compromise is called for. 
One reasonable choice would be $\thetadet = (1/2)\fwhm$, 
corresponding to a minimum detection $S/N\sim 3$, with $\qst\sim 3$ and
$\thetapix/\fwhm\sim 1/2$.  I henceforth adopt these values
for the following examples, which now completely specifies our
detection routine.   

        The dot--dashed lines in Figure 1 show the
resulting standard
mass detection curves.  They all 
lie above the optimal detection curves,
implying a lower overall sensitivity, as expected;
and as in the previous cases, they  
fall with $z$.  The low resolution
examples in Figure 1c show a slight
turn--down at low redshift, but under
no circumstances will they ever reach the
origin at $z=0$, as do the unresolved and
optimal resolved detection curves:  as 
already mentioned, there 
always remains a non--zero detection mass
at low $z$ in the standard case, due to the
surface brightness cut.  This constraint
may be neatly summarized, using our earlier
result, as $\yo[\MdetectSt(z=0),0]=\tildy/2\pi$ 
(but it must be noted that this relies on 
our use of the self--similar cluster model).
The loss of close--by, low--mass halos
is particularly noteworthy 
for the study of low--mass halos; to find them
in an SZ survey, one of the important potentials
of such efforts, will require special ``tuning'' of
detection criteria, to more closely approach the
optimal routine.  

\subsubsection{Counts}

        Our next goal is to use the detection
mass to calculate the cluster counts and 
redshift distributions.
It is worth noting in passing that one
can envision several different kinds of
source counts: as a function of the value
of $\thetadet$, as a function of some
aperture flux density (fixed or isophotal)
or as a function of survey sensitivity, 
$\sigpix$.  Only the last one, however, is useful
for optimizing an observing strategy, i.e.,
to answer the question of whether it is
better to `go deep', or to 'go wide' when
performing a survey.  We are thus brought
to consider in detail the detection mass as 
a function of detector sensitivity -- 
$\MdetectSt(z;\sigpix)$.  The most direct efficient
way to do this for a large number of 
different sensitivities is by
returning to Eq.(\ref{eq:thetadet_dimless}),
fixing $\thetadet (=1/2\fwhm)$ and then finding 
the detection mass as the root of the equation
for each $z$.  This avoids having to calculate
the entire $\thetadet$ surface for each
$\sigpix$ ($\tildy$) just to extract
a single contour.  

     Using the result of this operation
in Eq.(\ref{eq:counts}), we find the counts
and redshift distributions displayed in
Figures 2 and 3 as the dot--dashed lines.
Not surprisingly, the counts are
lower at a given sensitivity than the 
corresponding optimal counts, and they are also
slightly steeper.  Similar remarks apply
to the results of Figure 3.  All of this 
is easily understood from the loss of low--surface 
brightness objects relative to the optimal
routine.  The essential
conclusion concerning observation strategy
remains the same: down to low flux
densities, deeper integrations should
yield more sources.  

\section{Discussion}

        The effects of resolving clusters must
be properly modeled to understand
the capabilities of possible ground--based 
surveys, as is clear from, for example,
Figure 2: predicted counts are lower
and steeper for resolved clusters relative 
to hypothetical SZ point sources.  Besides
lowering the expectations for the number of
detectable sources, these results also
suggest that deep integrations are 
more efficient than wide and shallow
ones.  The actual number of clusters 
expected for realistic ground--based performance
are model dependent.  For a self--similar
cluster population, one could reasonably
expect between 10--100 clusters/sq. deg.
down to $0.1$ mJy at $2$ mm  and with $\fwhm=1$ arcmin,
as shown in Figure 2a.  This number 
depends in addition on the source detection method employed:
the standard routine counts may perhaps be
considered realistic, while the optimal method
counts indicate instead the best one could hope to 
achieve.  At $1$ cm, for equivalent sensitivity and at
the lower resolution of $\fwhm=2$ arcmin, one
expects an order of magnitude lower surface 
density (Figure 1c).   In sum, a square
degree survey at $2$ mm could yield 
$\sim 10-100$ detections depending on the
exact cluster model and the detection
algorithm; a 10 square degree survey
at $1$ cm to the same sensitivity ($0.1$ mJy)
could produce similar numbers.  Both types
of survey may soon be achievable, with instruments
similar to BOLOCAM (Glenn et al. 1998) or a detected interferometer
array (Carlstrom et al. 1999)

        One of the primary interests of opening 
this new window onto the Universe is to the search for
high redshift clusters.  The details of resolved 
cluster detection do not change the important
and tell--tail difference between the redshift
distributions in
different cosmological models:  the
expected number of high redshift clusters is
a sensitive function of $\Omo$,
as demonstrated by the redshift
distributions given in Figure 3.
Observations of such redshift distributions
should prove a valuable tool for constraining
$\Omo$ and for understanding evolution
of the cluster environment.

%    Figure 2a tells us that a survey with arcminute resolution
%down to $0.1$ mJy at 2 mm is susceptible 
%to finding between 10 to 100 objects/sq. deg.  
%The exact number depends on the cosmology/cluster model,
%and on the exact algorithm used to extract the
%clusters.  The standard detection counts may perhaps be considered
%more realistic than the optimal method counts, which indicate 
%instead the best one could achieve.  The smaller intrinsic SZ signal 
%at $1$ cm leads to a smaller number at equivalent sensitivity 
%and resolution.  This implies that, using the example of Figure 
%2c, a survey to the same limiting sensitivity, at the lower resolution 
%of $\fwhm\sim 2$ arcmins, could expect roughly an
%order of magnitude fewer sources.  I note
%that this is rather more pessimistic than
%the estimates given by Carlstrom et al. (1999),
%probably in part because it is founded on the
%standard detection routine.  The optimal 
%method in this case and for the open model
%predicts $\sim 10$ sources/sq. deg., closer
%to the numbers given by the previous authors.
%It is also partly due to the lower gas
%mass fraction used in the present work.  
%This latter point offers the occasion
%to re--emphasize the fact that the
%overall normalization of the counts
%is model dependent. 

        There are several important issues
that have not been dealt with in the 
present work.  One concerns 
eventual source confusion, an effect 
that depends on the beam size and
the exact value of the counts.
This effect may very well be important even
on arcminute scales, as noted by Aghanim et al. 
(1997).  As these authors also point out, 
the issue is complicated by the
fact that, due to the extended nature of
clusters, one must also contend with
{\em source blending}.  Detailed modeling
of these effects really requires simulations.

     Another important issue not addressed
in the present work concerns the question of
radio source contamination.  With sufficient
frequency coverage, on can always 
identify SZ sources by their unique 
spectrum.  Most often, though, spectral
coverage is limited and contamination
may become problematic.  Its importance
depends on the observation frequency, and
the counts at millimeter wavelengths
are in fact a subject of current fundamental
research; thus, the nature of contamination
at in the millimeter is much more model
dependent than in the centimeter.

     Finally, I note once again that the
present work is based on a simple cluster
model, because the principal motivation
has been to understand the nature of 
resolved cluster detection by comparison
to the more classic unresolved case.  
Any attempt at a more exact examination
of the number counts and redshift distributions
requires more detailed cluster modeling.
Such work would, in addition, permit
an interesting comparison of the 
relative efficiencies of SZ and X--ray 
observations to finding high redshift
clusters, {\em in practice}.  The 
SZ effect is clearly {\em inherently}
more efficient, but to really address
this question, one should consider
the actual achievable sensitivities
of the two approaches.  

\section{Conclusions}

        There are clear and important differences
in the conclusions one draws concerning 
SZ surveys depending on whether clusters
are considered as point sources or as extended.
For low resolution surveys, such as expected 
from the Planck Surveyor, most clusters will
remain unresolved; however, when discussing
the arcminute resolution more applicable to
possible future ground--based surveys,
we have seen that it is important to 
model the clusters as resolved sources 
in order to properly understand the 
nature of detectable objects.  For a
given sensitivity, high angular resolution
``resolves out'' some clusters, lowering
and steepening the final source counts.
Relative to optimal resolved detection, standard
algorithms tend to in addition loose
low mass, low redshift clusters due
to their imposed surface brightness cut,
further steepening and lowering the
counts.  With a fixed total observation
time and a given frequency and angular 
resolution, we have seen that our 
results imply that deep integrations
yield more objects than shallow ones
covering a large area.

     Some important issues still
to be explored concern the questions of
source confusion and blending, and radio
source contamination.  A detailed 
comparison of SZ and X--ray surveys
would also be of interest, which implies
more detailed cluster modeling than employed
here.  
        
        All the same, the numbers from the self--similar
cluster model should be, within all the present uncertainties
of these predictions, illustrative of what may be
soon achieved from the ground.  It appears that
both in the millimeter and in the cm, ground--based
SZ surveys could be capable of detecting up
to $\sim 100$ clusters in total, a respectable
statistical catalog.

\begin{acknowledgements}
I am very pleased to thank K.~Romer and J.~Mohr for their
SZ workshop at the centennial AAS meeting in Chicago, which
was the starting point for this work.  I am also grateful
for the hospitality of P.~Rosati at the European Southern
Observatory and of J.~Willick at Stanford University
where some of this work was carried out.  Thanks 
to B.~Keating for helpful discussions.     
\end{acknowledgements}

\end{document}